\documentclass[authoryear,review,11pt,a4paper]{elsarticle}

\usepackage{setspace}
\usepackage{fullpage}

\usepackage{color}

\usepackage{amsmath}

\usepackage{amsfonts}
\usepackage{amssymb}
\usepackage{amsthm}

\usepackage{amscd}

\usepackage{bbm}

\usepackage[ruled]{algorithm2e}

\usepackage{graphicx}

\usepackage{lscape}

\usepackage{longtable}
\usepackage{booktabs}
\usepackage{multirow}
\usepackage{float}
\usepackage{caption}
\usepackage{subcaption}

\usepackage{lineno}

\usepackage{hyperref}

\biboptions{authoryear}

\newtheorem{lem}{Lemma}
\newdefinition{rmk}{Remark}
\newproof{pf}{Proof}

\journal{ }

\begin{document}

\begin{frontmatter}



\title{Rapid evaluation and calibration of Bayesian group sequential designs via conjugate-mixture semi-simulation}

\author[TGIUK,ICTU]{Zhangyi He\corref{cor1}\fnref{fn3}}
\ead{zhe1@georgeinstitute.org.uk}


\author[BrisMath]{Feng Yu}

\author[ICTU]{Suzie Cro}

\author[TGIAUS]{Laurent Billot}

\cortext[cor1]{Corresponding author.}
\fntext[fn3]{Affiliation at time of study}

\address[TGIUK]{The George Institute for Global Health, Imperial College London, London W12 7RZ, United Kingdom}
\address[ICTU]{Imperial Clinical Trials Unit, Imperial College London, London W12 7RH, United Kingdom}
\address[BrisMath]{School of Mathematics, University of Bristol, Bristol BS8 1QU, United Kingdom}
\address[TGIAUS]{The George Institute for Global Health, University of New South Wales, Sydney 2042, Australia}

\begin{abstract}
Bayesian group sequential designs (GSDs) extend frequentist GSDs with interpretable decision-making and external evidence borrowing, but their use is limited by the computational burden of design-stage operating-characteristic evaluation. Conventional methods simulate virtual trials with Markov chain Monte Carlo or approximate analytical posterior updates at each interim look, making joint calibration of decision thresholds and design skeletons impractical on commodity hardware. Here we introduce a semi-simulation framework with two innovations. First, finite conjugate-mixture priors replace the posterior computation for each look (``per-look'') with closed-form conjugate updates and low-dimensional numerical integration for decision-rule tail probabilities. Second, a precomputation strategy caches per-look posterior tail probabilities from a single Monte Carlo pass at the union of all candidate analysis times, and each design in the calibration grid is evaluated against the same cache by a sub-second sweep, with no further simulation cost. The framework supports posterior-probability decision rules with multiple efficacy and futility criteria under either binding or non-binding futility, and derives closed-form per-look updates for binary, continuous, count and time-to-event endpoints, with benchmarking here focused on the binary endpoint. When applied to re-design the ADRENAL trial, with up to nine analyses, the framework reproduces the operating characteristics of \texttt{BATSS} and \texttt{adaptr} within Monte Carlo error while running, per GSD, approximately $7\times$ to $16\times$ faster than \texttt{adaptr} at a matched budget (several hundredfold at the million-trial calibration budget) and $3{,}700\times$ to $6{,}600\times$ faster than \texttt{BATSS}. 
This brings routine Bayesian GSD calibration within computational reach for confirmatory trials.
\end{abstract}


\end{frontmatter}


\section{Introduction}
\label{sec:1}
Phase III confirmatory randomised controlled trials are widely regarded as the gold standard for evaluating the efficacy and safety of new treatments in evidence-based medicine \citep{jennison2005,pallmann2018}. They can also be long and expensive, and over the last few decades, sustained interest has built around adaptive designs that can shorten or sharpen them without sacrificing the validity of the comparison. The most commonly adopted adaptive design for confirmatory trials is the group sequential design (GSD) \citep{stevely2015,judge2021}, which allows pre-planned interim analyses on accumulating data so that the trial can stop early when there is sufficient evidence of either efficacy or futility, while maintaining its integrity. Compared to fixed-sample designs, GSDs can deliver substantial savings in timeline, cost, and patient exposure. As reported in \citet{stevely2015}, the majority (around $68\%$) of confirmatory trials that adopted a (frequentist) GSD stopped early, predominantly for efficacy or futility.

Frequentist GSDs are well established \citep{jennison1999,wassmer2016} and commonly adopted \citep[\textit{e.g.},][]{keh2016,combes2018,perkins2018}. At each interim analysis, a test statistic is compared against critical values calibrated through an alpha-spending function so that the overall type I error rate is controlled \citep[\textit{e.g.},][]{haybittle1971,peto1976,pocock1977,obrien1979,lan1983,kim1987,hwang1990}. By contrast, their Bayesian counterparts have grown rapidly only over the past two decades \citep[see][and references therein]{zhou2024}, and the regulatory environment has matured to support their use in confirmatory settings, both for adaptive designs generally \citep{fda2019} and for Bayesian methodology specifically \citep[the latter currently in draft]{fda2010,fda2026}. In a Bayesian GSD, prior beliefs about the treatment effect are encoded in a prior distribution and are updated using accumulating trial data through Bayes' rule at each analysis, and decisions are made on quantities such as the posterior probability that the treatment is effective or the predictive probability of trial success \citep{spiegelhalter2004,berry2010,gsponer2014,saville2014,lee2024}. The Bayesian formulation supports the principled incorporation of historical or external evidence through borrowing \citep[\textit{e.g.},][]{schmidli2014,ibrahim2015,jiang2023,fda2023} and produces clinically interpretable decision criteria stated directly on the probability scale \citep{freedman1994,zhou2024}. 

Despite these advantages, Bayesian GSDs remain underused in confirmatory trials \citep{winkler2001,pibouleau2011,brard2017}. The principal obstacle is computational. Unlike spending functions, which give straightforward control of error rates in classical GSDs, Probability thresholds are typically calibrated through Monte Carlo simulation of virtual trials. When the posterior is not analytically tractable, each virtual trial requires Markov chain Monte Carlo (MCMC) at every look, with attendant convergence diagnostics. Nesting MCMC inside the trial-simulation loop quickly becomes prohibitive even for modest design grids \citep{lee2024}. Existing software addresses this computational barrier in two limited ways. The \texttt{R} package \texttt{gsbDesign} \citep{gerber2016} exploits the normal--normal conjugate pair to obtain closed-form per-look posterior updates, but it is restricted to normal priors with normal endpoints. The \texttt{R} package \texttt{adaptr} \citep{granholm2022} also uses conjugate-pair updating and is computationally efficient in the inner simulation loop, but its main focus is response-adaptive randomisation and arm-dropping in multi-arm trials, and its prior specification is restricted to the fixed conjugate families of its built-in models, with no mechanism to accommodate priors outside those families. In contrast, the \texttt{R} package \texttt{BATSS} \citep{couturier2024} replaces MCMC with the integrated nested Laplace approximation \citep[INLA;][]{rue2009}, allowing a broader range of priors and endpoints. This additional flexibility comes at a computational cost: \texttt{BATSS} remains considerably slower than methods based on direct conjugate updating (\textit{e.g.}, \texttt{gsbDesign} and \texttt{adaptr}). Consequently, no widely available tool for Bayesian GSDs currently combines the per-look speed of conjugate updating with broad flexibility in prior specification.

We address this gap by developing a semi-simulation framework built on two complementary ideas. First, we replace the per-look posterior computation with closed-form conjugate updates whenever the user-specified prior can be accurately approximated by a finite mixture of conjugate components. The posterior tail probabilities required by the decision rules then reduce to low-dimensional numerical integrals, eliminating the need for MCMC within the trial-simulation loop while retaining substantial flexibility in prior specification. Second, we introduce a precomputation strategy that separates trial simulation from design evaluation. A single Monte Carlo pass at the union of all candidate look times caches the per-look posterior tail probabilities at the candidate effect thresholds.
Every subsequent design in the calibration grid (regardless of the number and timing of interim analyses, the binding or non-binding futility convention, and the posterior-probability thresholds) is then evaluated against the same cache in milliseconds. Joint calibration of decision thresholds and the design skeleton against prespecified type I and type II error targets is therefore reduced to a sweep over cached quantities, rather than requiring a fresh simulation for each candidate design as in \texttt{BATSS} and \texttt{adaptr}.

Our implementation supports posterior-probability decision rules with multiple efficacy and futility criteria, allowing futility to be treated as either binding or non-binding. It also accommodates simple recruitment, follow-up, and dropout models such that expected study duration can be reported alongside the operating characteristics when required. The framework is deliberately scoped to fixed-allocation, single-comparison designs. Multiple comparisons, covariate adjustment and response-adaptive randomisation are outside its current scope and are discussed as directions for future work in Section~\ref{sec:4}. The closed-form per-look updates and precomputation strategy transform Bayesian GSD calibration from a computation-intensive off-line exercise into an interactive workflow feasible on a standard workstation, thereby lowering one of the practical barriers to routine use in confirmatory trials.

The remainder of this paper is organised as follows. Section~\ref{sec:2} develops the semi-simulation framework for Bayesian GSDs
and details the prior approximation, posterior computation, operating characteristics evaluation and the precomputation strategy that enables joint calibration of decision thresholds and the design skeleton. Section~\ref{sec:3} applies the framework to re-design the ADRENAL trial, benchmarks its performance against \texttt{BATSS} and \texttt{adaptr}, evaluates sensitivity to the number of interim looks and the choice of decision thresholds, presents calibrated three- and five-look designs, and assesses the Monte Carlo budget required for reliable calibration. Section~\ref{sec:4} discusses the contribution, limitations, software comparisons, and future extensions.

\section{Methods}
\label{sec:2}
We consider a two-arm group sequential trial with a binary primary outcome. Let $\vartheta_{c}$ be the response rate in the control arm and $\vartheta_{t}$ the response rate in the treatment arm. Our objective is to test the null hypothesis $H_{0}: \vartheta_{t}-\vartheta_{c}\leq 0$ against the one-sided alternative $H_{1}: \vartheta_{t}-\vartheta_{c}>0$, and to conclude that the treatment is superior to the control if $H_{0}$ is rejected. The trial includes a total of $K$ looks, made up of $K-1$ interim analyses and one final analysis. We present the development in this section for binary endpoints, with the treatment effect mainly parameterised on the absolute risk-difference scale and extended to the risk-ratio and odds-ratio scales, 
and extensions to continuous, count, and time-to-event endpoints using their natural conjugate pairs are provided in the Supplemental Material, Section~\href{run:./ZH2023_Supplemental_Material.pdf}{S1}.

\subsection{Decision criteria}
\label{sec:21}
Bayesian GSDs typically use decision rules based on posterior probabilities of the treatment effect, hereafter denoted PostPr. These rules determine stopping decisions at each interim and final analysis based on the posterior probability of the treatment effect reaching a prespecified critical value given the total amount of data available up to that point. Decision criteria can be independently specified at each look by combining requirements on the treatment effect and its associated posterior probability \citep{lee2024}. 

More specifically, we let $\pi_{k}(\delta \mid i,j)$ denote the posterior for the treatment effect $\Delta = \vartheta_{t}-\vartheta_{c}$ at the $k$-th analysis given the cumulative number of responses in each arm, $I_{k}=i$ and $J_{k}=j$. At the $k$-th interim analysis, we stop the trial for early efficacy if
\begin{linenomath}
	\begin{equation}
		\label{eqn:211}
		\mathbb{P}(\Delta>\delta^{E}_{k,u} \mid I_{k}=i,J_{k}=j)
		=
		\int_{\delta^{E}_{k,u}}^{1} \pi_{k}(\delta \mid i,j) \, d\delta
		>
		\gamma^{E}_{k,u}
	\end{equation}
\end{linenomath}
for $u=1,2,\ldots,U_{k}$, where $U_{k}$ is the number of efficacy criteria at the $k$-th analysis, and $\delta^{E}_{k,u}$ and $\gamma^{E}_{k,u}$ are the specified treatment-effect threshold and posterior-probability threshold for the $u$-th efficacy criterion. For example, \citet{gsponer2014} recommended dual efficacy criteria requiring evidence for both any positive treatment effect and a clinically meaningful effect.  We stop the trial for early futility if
\begin{linenomath}
	\begin{equation}
		\label{eqn:212}
		\mathbb{P}(\Delta<\delta^{F}_{k,v} \mid I_{k}=i,J_{k}=j)
		=
		\int_{-1}^{\delta^{F}_{k,v}} \pi_{k}(\delta \mid i,j) \, d\delta
		>
		\gamma^{F}_{k,v}
	\end{equation}
\end{linenomath}
for $v=1,2,\ldots,V_{k}$, where $V_{k}$ is the number of futility criteria at the $k$-th analysis, and $\delta^{F}_{k,v}$ and $\gamma^{F}_{k,v}$ are the specified treatment-effect threshold and posterior-probability threshold for the $v$-th futility criterion. Otherwise, we continue the trial. 
For simplicity, we introduce
\begin{linenomath}
	\begin{equation}
		\label{eqn:213}
		C_{k}
		=
		\prod_{u=1}^{U_{k}}\mathbbm{1}_{\{\mathbb{P}(\Delta>\delta^{E}_{k,u} \mid I_{k}=i,J_{k}=j)>\gamma^{E}_{k,u}\}} - \prod_{v=1}^{V_{k}}\mathbbm{1}_{\{\mathbb{P}(\Delta<\delta^{F}_{k,v} \mid I_{k}=i,J_{k}=j)>\gamma^{F}_{k,v}\}}
	\end{equation}
\end{linenomath}
to represent the result of the trial at the $k$-th interim analysis, where $\mathbbm{1}_{A}$ denotes the indicator function, equal to 1 when event $A$ occurs and 0 otherwise. The trial stops and declares efficacy if $C_{k}=1$ , stops and declares futility if $C_{k}=-1$, and otherwise continues. 
The products over $u$ and $v$ in Eq.~(\ref{eqn:213}) impose a logical AND over the efficacy and futility criteria, respectively, so that efficacy stopping requires all efficacy criteria to be satisfied and futility stopping requires all futility criteria to be met. Although $C_{k}=0$ also when both products equal 1, under routine specifications this tie cannot occur, because the efficacy events $\{\Delta>\delta^{E}_{k,u}\}$ and futility events $\{\Delta<\delta^{F}_{k,v}\}$ are typically disjoint. 
If simultaneous efficacy and futility triggers are possible, the rule can be modified to give priority to either efficacy or futility. A logical OR over criteria can be implemented by treating each disjunct as a separate stopping rule and stopping when any component rule is satisfied.

At the final analysis, the decision rules for efficacy, as specified in Eq.~(\ref{eqn:211}), are typically used to determine whether to reject $H_{0}$ or not, leading to 
\begin{linenomath}
	\begin{equation}
		\label{eqn:214}
		C_{K}
		=
		2\prod_{u=1}^{U_{K}}\mathbbm{1}_{\{\mathbb{P}(\Delta>\delta^{E}_{K,u} \mid I_{K}=i,J_{K}=j)>\gamma^{E}_{K,u}\}}-1.
	\end{equation}
\end{linenomath}
This implies a binary trial result at the final analysis, with the trial declaring efficacy if $C_{K}=1$ and otherwise terminating without an efficacy conclusion. We retain the label $C_{K}=-1$ for the latter outcome, for notational symmetry with the interim stages, but emphasise that ``futility'' at the final analysis means failure to reject $H_{0}$ rather than a separate futility stopping rule.

\subsection{Posterior computation}
\label{sec:22}
To compute the posterior $\pi_{k}(\delta \mid i,j)$ in Eqs.~(\ref{eqn:211}) and (\ref{eqn:212}), appropriate priors for $\vartheta_{c}$ and $\vartheta_{t}$ are required. These priors are commonly derived from historical data and/or clinical opinion, if available. Here we assume Beta priors for $\vartheta_{c}$ and $\vartheta_{t}$, \textit{i.e.}, $\vartheta_{c} \sim \operatorname{Beta}(a_{c},b_{c})$ and $\vartheta_{t} \sim \operatorname{Beta}(a_{t},b_{t})$, respectively, with the probability density function for $\operatorname{Beta}(a,b)$ given by
\begin{linenomath}
	\begin{equation*}
		f(\vartheta;a,b)
		=
		\frac{\vartheta^{a-1}(1-\vartheta)^{b-1}}{B(a,b)},
	\end{equation*}
\end{linenomath}
where $B(a,b)$ is the Beta function. General prior specifications will be discussed in Section~\ref{sec:23}.

At the $k$-th analysis, let $m_{k}$ and $n_{k}$ denote the cumulative numbers of subjects with available data in the control and treatment arms, respectively. Since the outcome of each subject independently and identically follows a Bernoulli distribution in each arm, the cumulative number of responses in each arm is binomially distributed, \textit{i.e.}, $I_{k} \sim \operatorname{Bin}(m_{k},\vartheta_{c})$ and $J_{k} \sim \operatorname{Bin}(n_{k},\vartheta_{t})$, respectively. 
We assume that $\vartheta_{c}$ and $\vartheta_{t}$ are a priori independent, so that the joint prior factorises as $\pi_{0}(\vartheta_{c},\vartheta_{t})=\pi_{0}(\vartheta_{c})\pi_{0}(\vartheta_{t})$ and posterior independence is preserved given the per-arm data. Under this assumption, the conjugate nature of the Beta prior combined with the binomial likelihood gives posteriors that remain Beta distributions, \textit{i.e.}, $\vartheta_{c} \mid I_{k}=i \sim \operatorname{Beta}(a_{c}+i,b_{c}+m_{k}-i)$ and $\vartheta_{t} \mid J_{k}=j \sim \operatorname{Beta}(a_{t}+j,b_{t}+n_{k}-j)$, respectively. We then have
\begin{linenomath}
	\begin{align}
			\label{eqn:221}
			\pi_{k}(\delta \mid i,j)
			&=
			\frac{d}{d\delta}\mathbb{P}(\Delta < \delta \mid I_{k}=i,J_{k}=j) \notag \\
			&=
			\frac{d}{d\delta}\int_{0}^{1}\int_{0}^{\min\{1,\max\{0,\vartheta_{c}+\delta\}\}} f(\vartheta_{c};a_{c}+i,b_{c}+m_{k}-i)f(\vartheta_{t};a_{t}+j,b_{t}+n_{k}-j) \, d\vartheta_{t}d\vartheta_{c} \notag \\
			&=
			\int_{\max\{0,-\delta\}}^{\min\{1,1-\delta\}} f(\vartheta;a_{c}+i,b_{c}+m_{k}-i)\,f(\vartheta+\delta;a_{t}+j,b_{t}+n_{k}-j) \, d\vartheta \notag \\
			&\propto
			\int_{\max\{0,-\delta\}}^{\min\{1,1-\delta\}} \vartheta^{a_{c}+i-1}(1-\vartheta)^{b_{c}+m_{k}-i-1}(\vartheta+\delta)^{a_{t}+j-1}(1-\vartheta-\delta)^{b_{t}+n_{k}-j-1} \, d\vartheta,
	\end{align}
\end{linenomath}
The posterior tail probabilities $\mathbb{P}(\Delta>\delta^{E}_{k,u} \mid I_{k}=i,J_{k}=j)$ and $\mathbb{P}(\Delta<\delta^{F}_{k,v} \mid I_{k}=i,J_{k}=j)$ in Eqs.~(\ref{eqn:211}) and (\ref{eqn:212}) are obtained by numerical integration of the density obtained from Eq.~(\ref{eqn:221})
over the corresponding subset of $[-1,1]$.

The same construction adapts to relative-effect parameterisations of $\Delta$ without altering the underlying conjugate update. For the risk ratio $\Delta=\vartheta_{t}/\vartheta_{c}$, the change of variables from $(\vartheta_{c},\vartheta_{t})$ to $(\vartheta_{c},\delta)$ has Jacobian $\vartheta_{c}$, which gives the posterior density function
\begin{linenomath}
	\begin{equation}
		\label{eqn:222}
		\pi_{k}(\delta \mid i,j)
		=
		\int_{0}^{\min\{1,1/\delta\}} \vartheta\,f(\vartheta;a_{c}+i,b_{c}+m_{k}-i)\,f(\vartheta\delta;a_{t}+j,b_{t}+n_{k}-j)\,d\vartheta.
	\end{equation}
\end{linenomath}
For the odds ratio $\Delta=[\vartheta_{t}/(1-\vartheta_{t})]/[\vartheta_{c}/(1-\vartheta_{c})]$, the change of variables from $(\vartheta_{c},\vartheta_{t})$ to $(\vartheta_{c},\delta)$ has Jacobian $\vartheta_{c}(1-\vartheta_{c})/(1-\vartheta_{c}+\vartheta_{c}\delta)^{2}$, which gives the posterior density function
\begin{linenomath}
	\begin{equation}
		\label{eqn:223}
		\pi_{k}(\delta \mid i,j)
		=
		\int_{0}^{1} \frac{\vartheta(1-\vartheta)}{(1-\vartheta+\vartheta\delta)^{2}}\,f(\vartheta;a_{c}+i,b_{c}+m_{k}-i)\,f\!\left(\frac{\vartheta\delta}{1-\vartheta+\vartheta\delta};a_{t}+j,b_{t}+n_{k}-j\right)\,d\vartheta.
	\end{equation}
\end{linenomath}
Both integrals are one-dimensional and are evaluated by the same numerical integration routine as Eq.~(\ref{eqn:221}). The corresponding posterior tail probabilities on the chosen relative effect scale are obtained in the same way as on the absolute scale.

\subsection{Prior approximation}
\label{sec:23}
The efficiency of our posterior computation in Bayesian GSDs with binary endpoints stems from the conjugate nature of the Beta prior when combined with the binomial likelihood. This conjugacy removes the need for the MCMC simulations required by conventional approaches \citep{lee2024} and so substantially reduces the computational burden.

To maintain this computational advantage for an arbitrary prior $\pi_{0}(\vartheta)$, we approximate the prior through a mixture of Beta distributions
\begin{linenomath}
	\begin{equation}
		\label{eqn:231}
		\hat{\pi}_{0}(\vartheta)
		=
		\sum_{l=1}^{L}w_{l}f(\vartheta;a_{l},b_{l})
	\end{equation}
\end{linenomath}
with $w_{l}>0$ for $l=1,2,\ldots,L$ and $\sum_{l=1}^{L}w_{l}=1$, where $L$ is the number of mixture components.
For each arm, the Beta-mixture approximation allows the posterior of the response rate to be expressed as a mixture of Beta distributions. 
This construction preserves conjugacy within our framework and so enables explicit and efficient posterior updates, even when the prior itself is non-conjugate. 
\citet{dalal1983} demonstrated that any prior on a bounded parameter space can be approximated arbitrarily well by a finite mixture of conjugate priors.

To construct the Beta-mixture approximation we need to determine the number of mixture components $L$, the mixture weights $\boldsymbol{w}_{1:L}=(w_{1},w_{2},\ldots,w_{L})^\intercal$, and the shape parameters $\boldsymbol{a}_{1:L}=(a_{1},a_{2},\ldots,a_{L})^\intercal$ and $\boldsymbol{b}_{1:L}=(b_{1},b_{2},\ldots,b_{L})^\intercal$, such that the approximate prior $\hat{\pi}_{0}(\vartheta)$ is sufficiently close to the exact prior $\pi_{0}(\vartheta)$. We use the forward Kullback--Leibler (KL) divergence \citep{kullback1951} of $\hat{\pi}_{0}(\vartheta)$ from $\pi_{0}(\vartheta)$, defined as
\begin{linenomath}
	\begin{equation}
		\label{eqn:233}
		D_{KL}(\pi_{0}(\vartheta) \parallel \hat{\pi}_{0}(\vartheta))
		=
		\int_{0}^{1}\pi_{0}(\vartheta)\log\left(\frac{\pi_{0}(\vartheta)}{\hat{\pi}_{0}(\vartheta)}\right) \,d\vartheta,
	\end{equation}
\end{linenomath}
to measure the divergence between them. The forward direction is mass-covering. It penalises $\hat{\pi}_{0}\to 0$ in regions where $\pi_{0}>0$ and so ensures that the approximation does not assign vanishing probability to regions of the parameter space supported by the elicited prior. This property matters in the present application, in which the posterior probability of small or large effects drives the stopping decision. With independent and identically distributed samples drawn from $\pi_{0}(\vartheta)$, minimising the empirical forward KL divergence is equivalent to maximum likelihood estimation of the mixture parameters and can be carried out through a standard expectation--maximisation (EM) algorithm \citep{dempster1977}.
For each specified $L$, the Beta-mixture approximation is obtained by selecting $\hat{\boldsymbol{w}}_{1:L}$, $\hat{\boldsymbol{a}}_{1:L}$, and $\hat{\boldsymbol{b}}_{1:L}$ to minimise $D_{KL}(\pi_{0}(\vartheta)\,\|\,\hat{\pi}_{0}(\vartheta))$ subject to $a_{l},b_{l},w_{l}>0$ for $l=1,2,\ldots,L$ and $\sum_{l=1}^{L}w_{l}=1$.

We fit a Beta mixture for each candidate size $L\in\{1,\ldots,L_{\max}\}$ and select $\hat{L}$ as the smallest mixture size whose empirical forward KL divergence lies within a prespecified tolerance $\varepsilon$ of the best value achieved across all candidate sizes. Specifically, writing $D_{L}=D_{KL}(\pi_{0}\parallel\hat{\pi}_{0}^{L})$ and $D_{\min}=\min_{1\leq L'\leq L_{\max}}D_{L'}$, we choose $\hat{L}$ as the smallest $L\in\{1,\ldots,L_{\max}\}$ such that $D_{L}-D_{\min}<\varepsilon$. This within-tolerance rule is well-defined even when the empirical KL is non-monotone in $L$ because of EM stochasticity, as observed in~\ref{sec:A1}, since the minimiser itself always meets the criterion. It also ensures that the selected mixture is no more complex than necessary while remaining close to the best available fit. 
The Bayesian information criterion provides an alternative selection rule that explicitly penalises additional components and can also be applied. A similar parsimony-driven procedure has been used by \citet{schmidli2014} in mixture-prior construction.

The achieved KL divergence also offers a useful prior-level diagnostic.
By Pinsker's inequality \citep[Lemma~2.5]{tsybakov2009},
\begin{linenomath}
	\begin{equation}
		\label{eqn:234}
		\sup_{A}|\mathbb{P}_{\pi_{0}}(A)-\mathbb{P}_{\hat{\pi}_{0}^{L}}(A)|
		\leq
		\sqrt{D_{KL}(\pi_{0}\parallel \hat{\pi}_{0}^{L})/2},
	\end{equation}
\end{linenomath}
where the supremum runs over all measurable events $A$, and $\mathbb{P}_{\pi_{0}}$ and $\mathbb{P}_{\hat{\pi}_{0}^{L}}$ represent probabilities computed under the target prior $\pi_{0}$ and its $L$-component Beta-mixture approximation $\hat{\pi}_{0}^{L}$, respectively, such that the left-hand side is the total-variation distance between $\pi_{0}$ and $\hat{\pi}_{0}^{L}$. 
This prior-level bound does not generally imply a uniform bound on the posterior tail probabilities entering the decision rules, but the empirical agreement between target-prior and approximate-prior tail probabilities at the decision-rule thresholds is substantially tighter than the prior-level worst-case bound (see~\ref{sec:A1}).

This approach also covers the effect-and-control specification, in which priors are assigned to the control rate $\vartheta_{c}$ and the treatment effect $\delta$. For $s=1,2,\ldots,S$, we draw $\vartheta_{c}^{(s)}\sim\pi_{0}(\vartheta_{c})$ and $\delta^{(s)}\sim\pi_{0}(\delta)$, set $\vartheta_{t}^{(s)}=h^{-1}(\vartheta_{c}^{(s)},\delta^{(s)})$, where $\Delta=h(\vartheta_{c},\vartheta_{t})$ is the treatment-effect contrast of Section~\ref{sec:22}, so that $h^{-1}(\vartheta_{c},\delta)$ returns the treatment rate consistent with an effect $\delta$ at control rate $\vartheta_{c}$. We then fit the Beta mixture in Eq.~(\ref{eqn:231}) separately to $\{\vartheta_{c}^{(s)}\}$ and $\{\vartheta_{t}^{(s)}\}$, which reduces the specification to the arm-specific setting. Separate marginal fitting discards the between-arm dependence induced by the transformation and restores the \textit{a priori} independence assumed in Section~\ref{sec:22}. This has little effect on the operating characteristics: the prior contributes only a small, fixed amount of information that is quickly outweighed by the accumulating trial data, so discarded features such as the between-arm dependence have diminishing influence on stopping decisions as information accrues, consistent with the Bernstein--von Mises phenomenon \citep[Theorem~10.1]{vandervaart1998}.

\subsection{Operating characteristics evaluation}
\label{sec:24}
Although Bayesian GSDs are structured around Bayesian principles, investigating their frequentist operating characteristics is essential, particularly in confirmatory trials, as this is what allows them to meet established regulatory standards for reliability, transparency, and efficiency \citep{fda2010,fda2026}. The operating characteristics of interest are the overall type I and type II error rates, the expected sample size and study duration, and the probabilities of stopping for efficacy or futility at each interim analysis under both the null and alternative hypotheses.

Monte Carlo simulation is typically used to estimate these characteristics by generating a large number of virtual trials, with the posterior computation at each look commonly carried out using MCMC techniques such as the Gibbs sampler or the Metropolis--Hastings algorithm \citep{lee2024}. This conventional nested simulation procedure is computationally burdensome for evaluating a single candidate design and becomes substantially more demanding when calibration requires searching across many candidate designs to meet prespecified targets.
We address both within our framework for Bayesian GSDs with PostPr rules, first deriving a rapid evaluation of the operating characteristics of any single design from closed-form conjugate posterior updates and low-dimensional numerical integration for the corresponding tail probabilities, and then exploiting a precomputation procedure that makes the evaluation cost almost fixed across candidate designs and so reduces calibration to a cache sweep.

As is the conventional approach, our procedure estimates operating characteristics by Monte Carlo simulation of virtual trials, but it analyses each simulated trial through the closed-form conjugate posterior updates of Section~\ref{sec:22} and the low-dimensional numerical integration of the corresponding tail probabilities. Let $\tau=\min\{k \in \{1,2,\ldots,K\} : C_{k} \neq 0\}$ denote the analysis at which the trial stops, and let $C$ represent the corresponding trial result. The realisations of $\tau$ and $C$ in the Monte Carlo simulations are written as $\tau^{(r)}$ and $C^{(r)}$ for $r=1,2,\ldots,R$, where $r$ indexes the virtual trial and $R$ is the total number of virtual trials. For data-generating values $\vartheta_{c}=\vartheta$ and $\vartheta_{t}=\vartheta+\delta$, we estimate the probabilities of stopping for efficacy and futility at each analysis by
\begin{linenomath}
	\begin{align}
		\label{eqn:241}
		\alpha_{1}^{\vartheta,\delta}
		&=
		\mathbb{P}_{\vartheta,\delta}(C_{1}=1)
		\approx
		\frac{1}{R}\sum_{r=1}^{R} \mathbbm{1}_{\{\tau^{(r)}=1,C^{(r)}=1 \mid \vartheta_{c}=\vartheta,\vartheta_{t}=\vartheta+\delta\}} \\
		\label{eqn:242}
		\alpha_{k}^{\vartheta,\delta}
		&=
		\mathbb{P}_{\vartheta,\delta}\left(\{C_{k}=1\} \cap \bigcap_{k'=1}^{k-1} \{C_{k'}=0\}\right)
		\approx
		\frac{1}{R}\sum_{r=1}^{R} \mathbbm{1}_{\{\tau^{(r)}=k,C^{(r)}=1 \mid \vartheta_{c}=\vartheta,\vartheta_{t}=\vartheta+\delta\}}
	\end{align}
\end{linenomath}
for $k=2,3,\ldots,K$ and
\begin{linenomath}
	\begin{align}
		\label{eqn:243}
		\beta_{1}^{\vartheta,\delta}
		&=
		\mathbb{P}_{\vartheta,\delta}(C_{1}=-1)
		\approx
		\frac{1}{R}\sum_{r=1}^{R} \mathbbm{1}_{\{\tau^{(r)}=1,C^{(r)}=-1 \mid \vartheta_{c}=\vartheta,\vartheta_{t}=\vartheta+\delta\}} \\
		\label{eqn:244}
		\beta_{k}^{\vartheta,\delta}
		&=
		\mathbb{P}_{\vartheta,\delta}\left(\{C_{k}=-1\} \cap \bigcap_{k'=1}^{k-1} \{C_{k'}=0\}\right)
		\approx
		\frac{1}{R}\sum_{r=1}^{R} \mathbbm{1}_{\{\tau^{(r)}=k,C^{(r)}=-1 \mid \vartheta_{c}=\vartheta,\vartheta_{t}=\vartheta+\delta\}}
	\end{align}
\end{linenomath}
for $k=2,3,\ldots,K$, where $\mathbb{P}_{\vartheta,\delta}$ represents the probability conditional on $\vartheta_{c}=\vartheta$ and $\vartheta_{t}=\vartheta+\delta$. These estimates converge to the true probabilities as the number of virtual trials increases.

We define $N$ to be the sample size and $T$ the study duration, respectively, and their expectations can be estimated with Eqs.~(\ref{eqn:241})--(\ref{eqn:244}) as 
\begin{linenomath}
	\begin{align}
		\label{eqn:245}
		\mathbb{E}_{\vartheta,\delta}(N)
		&=
		\sum_{k=1}^{K}\mathbb{E}_{\vartheta,\delta}(N \mid \tau=k)(\alpha_{k}^{\vartheta,\delta}+\beta_{k}^{\vartheta,\delta}) \notag \\
		&\approx
		\sum_{k=1}^{K}(m_{k}+n_{k})\frac{1}{R}\sum_{r=1}^{R} \mathbbm{1}_{\{\tau^{(r)}=k \mid \vartheta_{c}=\vartheta,\vartheta_{t}=\vartheta+\delta\}} \\
		\label{eqn:246}
		\mathbb{E}_{\vartheta,\delta}(T)
		&=
		\sum_{k=1}^{K}\mathbb{E}_{\vartheta,\delta}(T \mid \tau=k)(\alpha_{k}^{\vartheta,\delta}+\beta_{k}^{\vartheta,\delta}) \notag \\
		&\approx
		\sum_{k=1}^{K}\mathbb{E}_{\vartheta,\delta}(T \mid N=m_{k}+n_{k})\frac{1}{R}\sum_{r=1}^{R} \mathbbm{1}_{\{\tau^{(r)}=k \mid \vartheta_{c}=\vartheta,\vartheta_{t}=\vartheta+\delta\}}, 
	\end{align}
\end{linenomath}
where $\mathbb{E}_{\vartheta,\delta}$ denotes the expectation conditional on $\vartheta_{c}=\vartheta$ and $\vartheta_{t}=\vartheta+\delta$. The step from the first to the second line of Eq.~(\ref{eqn:245}) uses $\mathbb{E}_{\vartheta,\delta}(N\mid \tau=k)=m_{k}+n_{k}$, which holds under the standard design-stage assumption that recruitment pauses at each look until the data are available. For endpoints with delayed outcome ascertainment where recruitment continues during the analysis window (\textit{e.g.}, 90-day mortality with rolling enrolment), the realised sample size at $\tau=k$ may exceed $m_{k}+n_{k}$, and Eq.~(\ref{eqn:245}) should be interpreted as the planned rather than the operational expected sample size. In Eq.~(\ref{eqn:246}), $\mathbb{E}_{\vartheta,\delta}(T \mid N=m_{k}+n_{k})$ can be calculated either analytically or numerically, depending on the patient recruitment and follow-up models specified. For endpoints with delayed assessment (\textit{e.g.}, 90-day mortality), $T$ is the time at which the primary outcome of the $(m_{k}+n_{k})$-th evaluable patient becomes available, so the recruitment time and the follow-up time both contribute. As a simple example, if the patient recruitment process is modelled as a Poisson point process with a constant rate $\lambda$ (patients per day) and the per-patient follow-up duration is a fixed constant $t_{f}$, then $\mathbb{E}_{\vartheta,\delta}(T \mid N=m_{k}+n_{k})=(m_{k}+n_{k})/\lambda + t_{f}$. Setting $t_{f}=0$ recovers the recruitment-only case.

The overall type I error rate, $\alpha$, can be written down as the sum of $\alpha_{k}^{\vartheta,0}$, the type I error spent at each analysis under the null hypothesis of $\delta=0$. We can then estimate the overall type I error rate with Eqs.~(\ref{eqn:241}) and (\ref{eqn:242}) as
\begin{linenomath}
	\begin{equation}
		\label{eqn:247}
		\alpha
		=
		\sum_{k=1}^{K}\alpha_{k}^{\vartheta,0}
		\approx
		\sum_{k=1}^{K}\frac{1}{R}\sum_{r=1}^{R} \mathbbm{1}_{\{\tau^{(r)}=k,C^{(r)}=1 \mid \vartheta_{c}=\vartheta,\vartheta_{t}=\vartheta\}}.
	\end{equation}
\end{linenomath}
Similarly, the overall type II error rate, $\beta$, can be decomposed into the sum of $\beta_{k}^{\vartheta,\delta^{*}}$, the type II error spent at each analysis under the alternative hypothesis of $\delta=\delta^{*}$, where $\delta^{*}$ is the assumed treatment effect. With Eqs.~(\ref{eqn:243}) and (\ref{eqn:244}), the overall type II error rate can be expressed as
\begin{linenomath}
	\begin{equation}
		\label{eqn:248}
		\beta
		=
		\sum_{k=1}^{K}\beta_{k}^{\vartheta,\delta^{*}}
		\approx
		\sum_{k=1}^{K}\frac{1}{R}\sum_{r=1}^{R} \mathbbm{1}_{\{\tau^{(r)}=k,C^{(r)}=-1 \mid \vartheta_{c}=\vartheta,\vartheta_{t}=\vartheta+\delta^{*}\}},
	\end{equation}
\end{linenomath}
therefore the power being $1-\beta$.

For clarity, we detail our procedure for evaluating the operating characteristics of Bayesian GSDs with PostPr rules in Algorithm~\ref{alg:24}, in which $\mathcal{C}$ denotes the set of treatment-effect thresholds $\{\boldsymbol{\delta}^{E}_{1:K,\,\cdot\,},\boldsymbol{\delta}^{F}_{1:K,\,\cdot\,}\}$ and posterior-probability thresholds $\{\boldsymbol{\gamma}^{E}_{1:K,\,\cdot\,},\boldsymbol{\gamma}^{F}_{1:K,\,\cdot\,}\}$ that specify the decision criteria, and $\mathcal{P}$ is the set of mixture weights and shape parameters that define the Beta-mixture prior approximation. The output vectors $\boldsymbol{\alpha}_{1:K}^{\vartheta,\delta}=(\alpha_{1}^{\vartheta,\delta},\ldots,\alpha_{K}^{\vartheta,\delta})^{\intercal}$ and $\boldsymbol{\beta}_{1:K}^{\vartheta,\delta}=(\beta_{1}^{\vartheta,\delta},\ldots,\beta_{K}^{\vartheta,\delta})^{\intercal}$ collect the per-look efficacy- and futility-stopping probabilities defined in Eqs.~(\ref{eqn:241})--(\ref{eqn:244}). The data-generating distribution is parameterised here as $(\vartheta_{c},\vartheta_{t})=(\vartheta,\vartheta+\delta)$, which corresponds to the absolute-risk scale. For relative-effect measures (risk ratio and odds ratio) the algorithm is unchanged except that $\vartheta_{t}$ is computed from $(\vartheta,\delta)$ via the relevant mapping in Section~\ref{sec:22}. The expected study duration can be calculated independently of Algorithm~\ref{alg:24}, even though the computation of $\mathbb{E}_{\vartheta,\delta}(T \mid N=m_{k}+n_{k})$ in Eq.~(\ref{eqn:246}) may itself require Monte Carlo simulation, depending on the recruitment and follow-up models.

\begin{algorithm}[!ht]
  \SetKwInOut{Input}{input}
  \SetKwInOut{Output}{output}
  \Input{$K,\boldsymbol{m}_{1:K},\boldsymbol{n}_{1:K},\mathcal{C},\mathcal{P},\vartheta,\delta,R$}
  \Output{$\boldsymbol{\alpha}_{1:K}^{\vartheta,\delta},\boldsymbol{\beta}_{1:K}^{\vartheta,\delta},\mathbb{E}_{\vartheta,\delta}(N)$}
  \Begin{
  	\tcp{Per-arm response rates derived from $(\vartheta,\delta)$ under the absolute risk parameterisation $(\vartheta_{c},\vartheta_{t})=(\vartheta,\vartheta+\delta)$, and for risk ratio or odds ratio scales, replace by the corresponding mapping from Section~\ref{sec:22}.}
  	set $\mathcal{R}=\{1,2,\ldots,R\}$\;
  	\For{$r \in \mathcal{R}$}{
  		draw $i_{1}^{(r)} \sim \operatorname{Bin}(m_{1},\vartheta)$ and $j_{1}^{(r)} \sim \operatorname{Bin}(n_{1},\vartheta+\delta)$\;
    	get $C^{(r)}$ through Eq.~(\ref{eqn:213}) with $i_{1}^{(r)},j_{1}^{(r)},m_{1},n_{1},\mathcal{C},\mathcal{P}$\;
  		get $\tau^{(r)}=1$ if $C^{(r)} \neq 0$ and $\tau^{(r)}=K$ otherwise\;
		}
		get $\alpha_{1}^{\vartheta,\delta}$ and $\beta_{1}^{\vartheta,\delta}$ through Eqs.~(\ref{eqn:241}) and (\ref{eqn:243}) with $\boldsymbol{\tau}^{(1:R)},\boldsymbol{C}^{(1:R)}$\;
		update $\mathcal{R}=\{r=1,2,\ldots,R \mid C^{(r)}=0\}$\;
    \For{$k \leftarrow 2$ \KwTo $K-1$}{
    	\For{$r \in \mathcal{R}$}{
    		draw $i_{k}^{(r)} \sim i_{k-1}^{(r)}+\operatorname{Bin}(m_{k}-m_{k-1},\vartheta)$ and $j_{k}^{(r)} \sim j_{k-1}^{(r)}+\operatorname{Bin}(n_{k}-n_{k-1},\vartheta+\delta)$\;
	    	update $C^{(r)}$ through Eq.~(\ref{eqn:213}) with $i_{k}^{(r)},j_{k}^{(r)},m_{k},n_{k},\mathcal{C},\mathcal{P}$\;
  			update $\tau^{(r)}=k$ if $C^{(r)} \neq 0$ and $\tau^{(r)}=K$ otherwise\;
  		}
  		get $\alpha_{k}^{\vartheta,\delta}$ and $\beta_{k}^{\vartheta,\delta}$ through Eqs.~(\ref{eqn:242}) and (\ref{eqn:244}) with $\boldsymbol{\tau}^{(1:R)},\boldsymbol{C}^{(1:R)}$\;
  		update $\mathcal{R}=\{r=1,2,\ldots,R \mid C^{(r)}=0\}$\;
    }
    \For{$r \in \mathcal{R}$}{
  		draw $i_{K}^{(r)} \sim i_{K-1}^{(r)}+\operatorname{Bin}(m_{K}-m_{K-1},\vartheta)$ and $j_{K}^{(r)} \sim j_{K-1}^{(r)}+\operatorname{Bin}(n_{K}-n_{K-1},\vartheta+\delta)$\;
    	update $C^{(r)}$ through Eq.~(\ref{eqn:214}) with $i_{K}^{(r)},j_{K}^{(r)},m_{K},n_{K},\mathcal{C},\mathcal{P}$\;
		}
		get $\alpha_{K}^{\vartheta,\delta}$ and $\beta_{K}^{\vartheta,\delta}$ through Eqs.~(\ref{eqn:242}) and (\ref{eqn:244}) with $\boldsymbol{\tau}^{(1:R)},\boldsymbol{C}^{(1:R)}$\;
    get $\mathbb{E}_{\vartheta,\delta}(N)$ through Eq.~(\ref{eqn:245}) with $\boldsymbol{m}_{1:K},\boldsymbol{n}_{1:K},\boldsymbol{\alpha}_{1:K}^{\vartheta,\delta},\boldsymbol{\beta}_{1:K}^{\vartheta,\delta}$\;
  }
  \caption{Procedure for evaluating operating characteristics of Bayesian GSDs.}
  \label{alg:24}
\end{algorithm}

In practice, identifying a Bayesian GSD that meets prespecified targets on the type I and type II error rates is a joint calibration problem in which the user explores operating characteristics across different numbers and timings of interim analyses, as well as a range of decision criteria. 
Our implementation reduces this search to a sweep over cached posterior tail probabilities. The reduction rests on the following invariance.

\begin{lem}[Cache invariance]
\label{lem:241}
Fix the prior and the per-look analysis model. For each simulated trial and each analysis index $k$, the per-look posterior tail probability $\mathbb{P}(\Delta>\delta \mid I_{k}=i,J_{k}=j)$ relies on the candidate design only through the cumulative event counts $(I_{k},J_{k})$ at look $k$ and the effect threshold $\delta$. It is therefore invariant to all decision thresholds, the binding or non-binding futility convention, and any refinement of the look schedule, provided each candidate analysis time is one of the simulated look times.
\end{lem}

The result is immediate. Under the finite-mixture-of-conjugates prior the posterior at look $k$ is a deterministic conjugate update of $(I_{k},J_{k})$ (Section~\ref{sec:22}), so its tail probability is a function of $(I_{k},J_{k},k,\delta)$ alone. The probability thresholds and the binding flag enter only downstream, when the cached tail probabilities are compared against $\gamma^{E}_{k,u}$ and $\gamma^{F}_{k,v}$ to determine stopping. The framework therefore precomputes the per-look posterior probabilities at all candidate effect thresholds in a one-off simulation pass and reuses the resulting cache for every candidate design in the search grid. Operating characteristics are then obtained by combining the cached posterior probabilities with the corresponding probability thresholds at each look. Calibration therefore reduces to a sweep over the cache, with no further simulation cost.

Two scope conditions delimit the cache's validity. First, Lemma~\ref{lem:241} requires every candidate analysis time to lie in the simulated union. Refining the schedule with an analysis time outside that union invalidates the affected looks and requires a fresh simulation pass. Second, when the look schedule is information- or event-driven rather than fixed in advance, as for the time-to-event endpoint, the union must be simulated on the same grid (calendar or enrolment time, or cumulative events) on which the candidate looks are defined. The time-to-event construction of Section~\href{run:./ZH2023_Supplemental_Material.pdf}{S1.3} makes this grid explicit.

The computational cost decomposes into a one-time precomputation component and a much smaller per-design evaluation component. The precomputation generates $R$ virtual trials and evaluates posterior probabilities at each candidate look and effect threshold. For each candidate design, evaluation then requires only applying the proposed stopping thresholds to the cached probabilities and aggregating the resulting trial-level outcomes.
We let $K_{\max}$ denote the largest number of looks across the candidate designs in the search grid. The one-time cost then scales with $R\,K_{\max}$ and with the size of the threshold grid, whereas the subsequent per-design cost is negligible. The cache stores one tail probability for each simulated trial, candidate look, and candidate effect threshold. Its memory footprint is therefore of order $R\,K_{\max}\,|\mathcal{E}|$ floating-point values, where $|\mathcal{E}|$ is the number of distinct candidate effect thresholds. 
The trial-level loop in Algorithm~\ref{alg:24} is embarrassingly parallel across $r=1,2,\ldots,R$, a structure that we exploit in the implementation. 
Implementation details and source code are available at \url{https://github.com/zhangyi-he/BayesGSD/tree/main/SemiSimul/PostPr}.

\section{Results}
\label{sec:3}
By following \citet{li2023}, we re-design the ADRENAL study \citep{venkatesh2018} under a Bayesian GSD framework, exploring different numbers and timings of interim analyses together with several alternative decision criteria, and evaluate their operating characteristics with our proposed approach. We validate the results by comparing them against those obtained from \texttt{adaptr} (version 1.5.0; \citealp{granholm2026}) and \texttt{BATSS} (version 1.1.1; \citealp{couturier2025}), and benchmark the computational efficiency of our framework against both packages. We then investigate the simulation size required to reliably evaluate the operating characteristics of Bayesian GSDs, with particular emphasis on accurate calculation of the overall type I error rate. All numerical results and wall-clock figures reported here are produced on a single workstation with a 4.2GHz quad-core Intel Core i7 CPU (8 logical threads through Hyper-Threading) and 32GB of 2400MHz DDR4 memory.

\subsection{Example study: the ADRENAL trial}
\label{sec:31}
The ADRENAL study was a multicentre, placebo-controlled, double-blind, randomised clinical trial designed to evaluate whether hydrocortisone, compared with placebo, reduces 90-day all-cause mortality among patients with septic shock admitted to the intensive care unit. The primary endpoint was all-cause mortality at 90 days after randomisation. Given a 33\% 90-day mortality rate in the placebo group, the study was powered to detect an absolute risk reduction of 5\% with hydrocortisone at a two-sided significance level of 0.05. The original planned sample size of 3{,}800 patients (1{,}900 per arm, assuming a 1\% dropout rate) provided 90\% power under these assumptions. The ADRENAL trial adopted a three-stage group sequential Haybittle--Peto design \citep{haybittle1971,peto1976} with two interim looks, conducted when primary endpoint data were available for 950 and 2{,}500 patients (corresponding to information fractions of 25.0\% and 65.8\%, respectively). Further details of the protocol and statistical analysis plan are available in \citet{venkatesh2018}.


\subsection{Bayesian re-design of the ADRENAL trial}
\label{sec:32}

The Bayesian re-design differs from the original frequentist ADRENAL design in three ways. First, the re-design is evaluated against the one-sided target $\alpha \leq 2.5\%$ throughout, corresponding to the original two-sided $0.05$ significance level under the one-sided alternative $H_{1}:\Delta<0$ adopted below. Second, the re-design includes posterior-probability rules for both efficacy and futility, rather than the efficacy-only Haybittle--Peto stopping rule used in the original protocol. This reflects the operational role that pre-specified futility monitoring now plays in modern confirmatory trials and aligns with the comparator implementations used in Section~\ref{sec:33}. Third, we use equally spaced look times, rather than the original information fractions of $25.0\%$ and $65.8\%$, to simplify the dependence of the operating characteristics on $K$ across the design grid. The precomputation strategy of Section~\ref{sec:24} accommodates arbitrary unequal look spacings at no additional cost, and could in principle search jointly over the look schedule and the threshold combinations, but we restrict to equal spacing here for clarity of exposition.

With these three departures fixed, we redesign the ADRENAL trial across $K\in\{1,2,\ldots,10\}$ total analyses. This includes a Bayesian fixed-sample design ($K=1$) and Bayesian GSDs with between one and nine equally spaced interim looks ($K\in\{2,3,\ldots,10\}$), each allowing stopping for either efficacy or futility. As the primary endpoint is mortality, clinical benefit corresponds to a reduction in the event rate, so the treatment effect is defined as $\Delta=\vartheta_t-\vartheta_c<0$. Accordingly, the null and alternative hypotheses are $H_{0}:\Delta\geq 0$ and $H_{1}:\Delta<0$. As clinical benefit lies in $\Delta<0$ rather than $\Delta>0$, the inequalities of Eqs.~(\ref{eqn:211})--(\ref{eqn:212}) flip throughout here, with the efficacy rule becoming $\mathbb{P}(\Delta<\delta_{k,u}^{E}\mid\,\cdot\,)>\gamma_{k,u}^{E}$ and the futility rule becoming $\mathbb{P}(\Delta>\delta_{k,v}^{F}\mid\,\cdot\,)>\gamma_{k,v}^{F}$. Algorithm~\ref{alg:24} applies unchanged once this sign-flip is propagated to the indicator functions of Eq.~(\ref{eqn:213}). 

We consider either a single efficacy criterion $\mathbb{P}(\Delta<0 \mid \,\cdot\,)>\gamma^{E}$ or a dual efficacy criterion, as in \citet{gsponer2014}, requiring both $\mathbb{P}(\Delta<0 \mid \,\cdot\,)>\gamma^{E}$ and $\mathbb{P}(\Delta<-0.05 \mid \,\cdot\,)>0.5$ at each look, where $\gamma^{E} \in \{0.90,0.95,0.99\}$ and the second criterion threshold of $0.5$ follows the convention of \citet{gsponer2014} that requires at least half of the posterior mass to lie beyond the clinically meaningful effect threshold. The futility criterion is $\mathbb{P}(\Delta>0 \mid \,\cdot\,)>\gamma^{F}$ at each interim look, with $\gamma^{F} \in \{0.85,0.90,0.95,1.00\}$. Setting $\gamma^{F}=1.00$ removes the futility rule, so that early stopping for futility is not permitted. Both binding and non-binding futility rules are considered.

For $K=1$, the futility threshold and the binding/non-binding distinction have no effect, so the grid collapses to $3$ efficacy thresholds $\times$ $2$ efficacy-criterion variants $= 6$ unique designs. For $K\in\{2,3,\ldots,10\}$, all five design axes are active: number of interim looks, efficacy threshold, futility threshold, binding/non-binding convention, and single versus dual efficacy criteria. These contribute $9 \times 3 \times 4 \times 2 \times 2 = 432$ configurations. Across these options, we thus evaluate $6+432=438$ distinct design configurations.

We set the maximum sample size to 3{,}800, matching the original ADRENAL trial design, and assess the operating characteristics of each candidate Bayesian GSD. The metrics of interest, namely the overall type I and type II error rates and the expected sample sizes, are produced through Algorithm~\ref{alg:24} based on $10^{6}$ simulated trials, with a uniform $\operatorname{Beta}(1,1)$ prior on both the treatment and control response rates. This prior is uniform on the response-rate scale and serves as a deliberately diffuse baseline with negligible information relative to the trial sample size. 
The type I error rate and the corresponding expected sample size are evaluated at $(\vartheta,\delta)=(0.33,0)$, representing the data-generating values under $H_{0}$. The type II error rate and the corresponding expected sample size are evaluated at $(\vartheta,\delta)=(0.33,-0.05)$, representing the data-generating values under the specified $H_{1}$. 
\subsection{Validation and benchmarking}
\label{sec:33}
We benchmark our framework against two reference implementations: \texttt{BATSS} (version 1.1.1; \citealp{couturier2025}), which uses \texttt{INLA} for posterior inference and supports a broad range of priors and endpoints, and \texttt{adaptr} (version 1.5.0; \citealp{granholm2026}), which adopts analytic Beta-binomial conjugate updates and is built mainly for response-adaptive randomisation but applies equally to fixed-allocation Bayesian GSDs. The package \texttt{gsbDesign} \citep{gerber2016} is the natural alternative for normal endpoints with normal priors but is not applicable to the binary endpoint of the ADRENAL re-design, so we do not include it in the head-to-head comparison. Section~\ref{sec:331} reports the benchmark under a uniform $\operatorname{Beta}(1,1)$ conjugate prior on the per-arm rates, under which the mixture-approximation step of Section~\ref{sec:23} collapses to a single component, and the comparison exercises the trial-simulation and decision-rule machinery alone. Section~\ref{sec:332} repeats the benchmark under an informative non-conjugate logit-normal prior on each arm, which assesses the mixture-approximation step and shows that the agreement between methods carries over under genuine prior information.

Beyond the prior distribution assigned to the arm-specific response rates, the two benchmark studies follow the same evaluation protocol. We use an efficacy threshold of $\gamma^{E}=0.99$, a futility threshold of $\gamma^{F}=0.90$, equal allocation, and a maximum sample size of $N=3{,}800$. We compare five designs: a fixed-sample design ($K=1$) and GSDs with $K \in \{3,5,7,9\}$ equally spaced looks at information fractions $j/K$ for $j=1,\ldots,K$. 

All three approaches use a single layer of \texttt{mclapply} parallelism, with one worker per logical CPU, and \texttt{INLA} is pinned to a single thread per worker (\texttt{num.threads}~$=$~``1:1'') so that \texttt{BATSS} and the proposed framework receive identical per-worker hardware budgets. To maintain computational tractability for the \texttt{adaptr} and \texttt{BATSS} runs while still enabling a fair head-to-head comparison, we run all three methods at a matched budget of $R=5{,}000$ simulated trials per scenario. 
We also run the proposed framework and \texttt{adaptr} at $R=10^{6}$, the routine high-precision budget we recommend for threshold calibration, to illustrate that this budget is readily feasible for both. 
\texttt{BATSS} is reported at the matched $R=5{,}000$ budget only because both its wall-clock and its memory footprint make $R=10^{6}$ simulations impractical on commodity hardware (\textit{e.g.}, the nine-look design under $H_{0}$ would require approximately 34 days and 500 GB memory, far exceeding the $32$ GB of the benchmark workstation). 

Neither \texttt{BATSS} nor \texttt{adaptr} caches its per-look posterior probabilities for reuse. In both, the threshold combination and the design skeleton are inputs to the simulator, so every candidate design in a calibration grid requires its own fresh simulation pass. Sections~\ref{sec:331} and~\ref{sec:332} thus report per-design wall-clock cost, but the practical headline of the framework is at the calibration stage, where the precomputation strategy of Section~\ref{sec:24} replaces a per-design simulation cost with a one-off simulation pass at the union of all candidate look times. Section~\ref{sec:34} returns to this point.

\subsubsection{Conjugate prior}
\label{sec:331}
The proposed framework and \texttt{adaptr} both assign independent $\operatorname{Beta}(1,1)$ priors directly to the arm-specific response rates. This prior is conjugate to the binomial likelihood in each arm, so the per-look update is closed-form, and the resulting per-arm posterior is Beta distributed. The mixture-approximation step of Section~\ref{sec:23} reduces to a single component and contributes no error. \texttt{BATSS} cannot accept the same prior because \texttt{INLA} parameterises the model on the logit scale. We therefore supply \texttt{BATSS} with \texttt{INLA}'s default weakly informative Gaussian priors on the regression coefficients, which provide the closest \texttt{INLA}-representable analogue. The comparison is therefore not exactly like-for-like at the prior level, but the difference is small relative to the binomial likelihood at $N=3{,}800$, and its empirical impact is assessed below.

The three approaches illustrate close agreement across all five designs and both hypotheses (Tables~\ref{tab:331} and~\ref{tab:332}). Taking the proposed framework at $R=10^{6}$ as a high-precision reference, with Monte Carlo standard error (SE) at most $0.02$ percentage points, the absolute difference in the estimated type I error rate at $R=5{,}000$ is at most $0.26$ percentage points for \texttt{BATSS} and $0.24$ percentage points for \texttt{adaptr}. The corresponding absolute differences in estimated power are at most $0.36$ percentage points for \texttt{BATSS} and $0.59$ percentage points for \texttt{adaptr}. All differences lie within the combined Monte Carlo SE of the two estimators being compared. Running \texttt{adaptr} at the same $R=10^{6}$ budget as for the proposed framework removes most of the Monte Carlo uncertainty from this comparison. At this precision, both methods have binomial SEs of about $0.02$ percentage points, and the two high-precision estimates agree to within $0.04$ percentage points on the type I error rate across all five designs and to within $0.08$ percentage points on power for every design with early stopping ($K\geq 3$), widening to $0.19$ percentage points only in the single-look case. Under this conjugate prior, the proposed framework's mixture reduces to a single exact Beta posterior ($\hat{L}=1$). Hence, this agreement confirms that its Gauss--Legendre evaluation of the posterior tail probabilities reproduces \texttt{adaptr}'s conjugate decision rule to Monte Carlo resolution rather than testing any prior approximation.

\begin{table}[!ht]
  \centering
  \begin{tabular}{rllrrrr}
    \toprule
    $K$ & Method & $R$ & Type I error & MC SE & $\mathbb{E}(N)$ & Time (s) \\
    \midrule
    $1$ & \texttt{BATSS} & $5{,}000$ & 0.92\,\% & 0.135\,pp & 3,800 & 2,101 \\
    $1$ & \texttt{adaptr} & $5{,}000$ & 1.22\,\% & 0.155\,pp & 3,800 & 9.9 \\
    $1$ & \texttt{adaptr} & $10^{6}$ & 1.02\,\% & 0.010\,pp & 3,800 & 2,064 \\
    $1$ & Proposed & $5{,}000$ & 0.98\,\% & 0.139\,pp & 3,800 & 0.3 \\
    $1$ & Proposed & $10^{6}$ & 0.98\,\% & 0.010\,pp & 3,800 & 1.6 \\
    \midrule
    $3$ & \texttt{BATSS} & $5{,}000$ & 2.22\,\% & 0.208\,pp & 3,436 & 5,920 \\
    $3$ & \texttt{adaptr} & $5{,}000$ & 2.44\,\% & 0.218\,pp & 3,445 & 13.1 \\
    $3$ & \texttt{adaptr} & $10^{6}$ & 2.29\,\% & 0.015\,pp & 3,446 & 2,606 \\
    $3$ & Proposed & $5{,}000$ & 2.26\,\% & 0.210\,pp & 3,448 & 0.9 \\
    $3$ & Proposed & $10^{6}$ & 2.30\,\% & 0.015\,pp & 3,443 & 3.5 \\
    \midrule
    $5$ & \texttt{BATSS} & $5{,}000$ & 2.90\,\% & 0.237\,pp & 3,231 & 8,756 \\
    $5$ & \texttt{adaptr} & $5{,}000$ & 3.26\,\% & 0.251\,pp & 3,238 & 16.5 \\
    $5$ & \texttt{adaptr} & $10^{6}$ & 3.09\,\% & 0.017\,pp & 3,245 & 3,514 \\
    $5$ & Proposed & $5{,}000$ & 3.04\,\% & 0.243\,pp & 3,260 & 1.6 \\
    $5$ & Proposed & $10^{6}$ & 3.11\,\% & 0.017\,pp & 3,242 & 6.0 \\
    \midrule
    $7$ & \texttt{BATSS} & $5{,}000$ & 3.46\,\% & 0.258\,pp & 3,118 & 12,153 \\
    $7$ & \texttt{adaptr} & $5{,}000$ & 3.86\,\% & 0.272\,pp & 3,120 & 20.8 \\
    $7$ & \texttt{adaptr} & $10^{6}$ & 3.74\,\% & 0.019\,pp & 3,106 & 3,754 \\
    $7$ & Proposed & $5{,}000$ & 3.98\,\% & 0.276\,pp & 3,128 & 1.9 \\
    $7$ & Proposed & $10^{6}$ & 3.72\,\% & 0.019\,pp & 3,108 & 8.4 \\
    \midrule
    $9$ & \texttt{BATSS} & $5{,}000$ & 4.28\,\% & 0.286\,pp & 3,007 & 14,619 \\
    $9$ & \texttt{adaptr} & $5{,}000$ & 4.26\,\% & 0.286\,pp & 3,009 & 21.5 \\
    $9$ & \texttt{adaptr} & $10^{6}$ & 4.19\,\% & 0.020\,pp & 3,002 & 4,428 \\
    $9$ & Proposed & $5{,}000$ & 3.70\,\% & 0.267\,pp & 2,994 & 2.3 \\
    $9$ & Proposed & $10^{6}$ & 4.15\,\% & 0.020\,pp & 2,999 & 9.6 \\
    \bottomrule
  \end{tabular}
  \caption{Type I error rate estimates. Estimates and wall-clock cost (in seconds) for the proposed semi-simulation framework, \texttt{BATSS} and \texttt{adaptr} on the fixed-sample design ($K=1$) and GSDs with three, five, seven and nine equally spaced looks ($K\in\{3,5,7,9\}$) for the ADRENAL re-design ($\vartheta=0.33$; the type I error rate and the corresponding $\mathbb{E}(N)$ are evaluated under $H_{0}$ at $\delta=0$; efficacy threshold $\gamma^{E}=0.99$, futility threshold $\gamma^{F}=0.90$, total sample size $N=3{,}800$). ``MC SE'' denotes the Monte Carlo SE of the corresponding estimate, in percentage points (pp). $\mathbb{E}(N)$ is the estimated expected total sample size. All three methods use independent $\operatorname{Beta}(1,1)$ priors on the per-arm rates, except that \texttt{BATSS} uses \texttt{INLA} with weakly informative Gaussian priors on the logit-scale coefficients.}
  \label{tab:331}
\end{table}
\begin{table}[!ht]
  \centering
  \begin{tabular}{rllrrrr}
    \toprule
    $K$ & Method & $R$ & Power & MC SE & $\mathbb{E}(N)$ & Time (s) \\
    \midrule
    $1$ & \texttt{BATSS} & $5{,}000$ & 84.32\,\% & 0.514\,pp & 3,800 & 2,102 \\
    $1$ & \texttt{adaptr} & $5{,}000$ & 84.64\,\% & 0.510\,pp & 3,800 & 12.6 \\
    $1$ & \texttt{adaptr} & $10^{6}$ & 84.49\,\% & 0.036\,pp & 3,800 & 2,296 \\
    $1$ & Proposed & $5{,}000$ & 84.32\,\% & 0.514\,pp & 3,800 & 0.3 \\
    $1$ & Proposed & $10^{6}$ & 84.68\,\% & 0.036\,pp & 3,800 & 1.3 \\
    \midrule
    $3$ & \texttt{BATSS} & $5{,}000$ & 86.64\,\% & 0.481\,pp & 2,498 & 4,007 \\
    $3$ & \texttt{adaptr} & $5{,}000$ & 86.88\,\% & 0.477\,pp & 2,511 & 13.2 \\
    $3$ & \texttt{adaptr} & $10^{6}$ & 86.75\,\% & 0.034\,pp & 2,497 & 2,639 \\
    $3$ & Proposed & $5{,}000$ & 85.88\,\% & 0.492\,pp & 2,482 & 0.8 \\
    $3$ & Proposed & $10^{6}$ & 86.79\,\% & 0.034\,pp & 2,494 & 3.7 \\
    \midrule
    $5$ & \texttt{BATSS} & $5{,}000$ & 87.46\,\% & 0.468\,pp & 2,205 & 6,313 \\
    $5$ & \texttt{adaptr} & $5{,}000$ & 87.36\,\% & 0.470\,pp & 2,207 & 16.2 \\
    $5$ & \texttt{adaptr} & $10^{6}$ & 87.69\,\% & 0.033\,pp & 2,206 & 2,962 \\
    $5$ & Proposed & $5{,}000$ & 87.76\,\% & 0.464\,pp & 2,203 & 1.4 \\
    $5$ & Proposed & $10^{6}$ & 87.77\,\% & 0.033\,pp & 2,202 & 5.2 \\
    \midrule
    $7$ & \texttt{BATSS} & $5{,}000$ & 88.56\,\% & 0.450\,pp & 2,017 & 7,871 \\
    $7$ & \texttt{adaptr} & $5{,}000$ & 88.56\,\% & 0.450\,pp & 2,048 & 15.9 \\
    $7$ & \texttt{adaptr} & $10^{6}$ & 88.18\,\% & 0.032\,pp & 2,054 & 3,260 \\
    $7$ & Proposed & $5{,}000$ & 88.94\,\% & 0.444\,pp & 2,033 & 2.1 \\
    $7$ & Proposed & $10^{6}$ & 88.24\,\% & 0.032\,pp & 2,052 & 6.7 \\
    \midrule
    $9$ & \texttt{BATSS} & $5{,}000$ & 88.54\,\% & 0.450\,pp & 1,929 & 9,534 \\
    $9$ & \texttt{adaptr} & $5{,}000$ & 89.02\,\% & 0.442\,pp & 1,942 & 18.4 \\
    $9$ & \texttt{adaptr} & $10^{6}$ & 88.45\,\% & 0.032\,pp & 1,956 & 3,569 \\
    $9$ & Proposed & $5{,}000$ & 88.24\,\% & 0.456\,pp & 1,951 & 2.2 \\
    $9$ & Proposed & $10^{6}$ & 88.43\,\% & 0.032\,pp & 1,954 & 7.9 \\
    \bottomrule
  \end{tabular}
  \caption{Power estimates. As Table~\ref{tab:331}, but reporting the estimated power rather than the type I error rate.}
  \label{tab:332}
\end{table}

To assess this empirically rather than depend on the binomial Monte Carlo SE alone, we re-run the proposed framework under $H_{0}$ at $R=5{,}000$ across $1{,}000$ independent seeds for the worst-case row of Table~\ref{tab:331} ($K=9$, $\gamma^{E}=0.99$, $\gamma^{F}=0.90$, with binding futility), which carries both the largest cross-method difference and the largest binomial Monte Carlo SE among the cells reported. The empirical seed-to-seed standard deviation (SD) is $0.277$ percentage points, closely matching the binomial Monte Carlo SE of $0.282$ percentage points at the empirical mean rate $p\approx 0.0416$ (the value $p=0.04$ gives $0.277$). The empirical mean is $4.16\%$, essentially identical to the $R=10^{6}$ reference estimate of $4.15\%$. The canonical-seed estimate reported in Table~\ref{tab:331}, $3.70\%$, lies $1.67$ empirical SDs below this mean, while the \texttt{BATSS} and \texttt{adaptr} estimates, $4.28\%$ and $4.26\%$, respectively, lie within $0.45$ empirical SDs. Thus, the matched-budget cross-method differences at $K=9$ are consistent with binomial Monte Carlo variation in the proposed-framework estimate, with no evidence of residual prior- or inference-induced bias in the comparator estimates beyond that margin.

The estimated expected sample sizes also agree closely. For every $K\geq 3$, the cell sits within $35$ patients of the proposed-framework reference (approximately $1\%$ of $N=3{,}800$), and the three approaches report $\mathbb{E}(N)=3{,}800$ identically for $K=1$ because no early stopping is possible. The remaining cross-method differences are consistent with Monte Carlo error, although they could in principle also reflect two methodological differences: the use of \texttt{INLA}'s logit-scale Gaussian priors in \texttt{BATSS} rather than the per-arm $\operatorname{Beta}(1,1)$ priors used by the other two methods, and the known tendency of \texttt{INLA} to underestimate posterior variances in models with binomial outcomes \citep{ferkingstad2015}. The benchmark cannot fully separate these two effects, but the empirical multi-seed check above bounds the magnitude of any residual systematic gap at well below the cross-method differences observed, and neither appears large enough at the thresholds considered here to alter design conclusions.

The three methods differ sharply in wall-clock cost per design, as shown in seconds in the last columns of Tables~\ref{tab:331} and~\ref{tab:332}. At the matched simulation budget of $R=5{,}000$ trials, across the four GSDs ($K\in\{3,5,7,9\}$), \texttt{BATSS} is roughly $3{,}700\times$ to $6{,}600\times$ slower than the proposed framework, reflecting the need for one \texttt{INLA} call per virtual trial at each look. \texttt{adaptr} is approximately $7\times$ to $16\times$ slower, reflecting the cost of its $5{,}000$ within-trial posterior draws on top of the conjugate Beta update, although it remains roughly $300\times$ to $680\times$ faster than \texttt{BATSS} at the same budget. The advantage over \texttt{adaptr} widens substantially at the high-precision budget. At $R=10^{6}$ the proposed framework completes each design in $1.3$ to $9.6$ seconds whereas \texttt{adaptr} takes $34$ to $74$ minutes (Tables~\ref{tab:331} and~\ref{tab:332}, last column), roughly $450\times$ to $740\times$ slower, since \texttt{adaptr}'s cost grows with its $5{,}000$ within-trial draws at every look across all $10^{6}$ simulated trials while the proposed framework's vectorised quadrature barely grows with $R$. \texttt{BATSS} is impractical at $R=10^{6}$, so even the proposed framework's $R=10^{6}$ runs remain about $1{,}100\times$ to $1{,}700\times$ faster than \texttt{BATSS} at the much smaller $R=5{,}000$ budget. These ratios characterise the per-design cost. The more consequential contrast is the cost of evaluating many candidate designs jointly. As illustrated in Section~\ref{sec:34}, the proposed precomputation strategy reduces the calibration of a $438$-design grid to an approximately seven-minute cache sweep on $8$ logical CPUs (about eight minutes end-to-end including the initial one-minute simulation pass at the union of look times), against a workload of roughly a CPU-month per hypothesis under \texttt{BATSS} on the same hardware.

\subsubsection{Non-conjugate prior}
\label{sec:332}
Whereas Section~\ref{sec:331} exercises the trial-simulation and decision-rule machinery under a conjugate prior, this section directly tests the mixture-approximation step. In this comparison, \texttt{adaptr} evaluates the logit-normal posterior nearly exactly by inverse-CDF sampling, without a Beta-mixture approximation. Agreement between \texttt{adaptr} and the proposed framework therefore provides a sharper check on the finite-mixture approximation than comparison with the proposed framework's own high-precision reference. We replace the $\operatorname{Beta}(1,1)$ prior of Section~\ref{sec:331} with an informative non-conjugate logit-normal prior on each arm, $\operatorname{logit}(\vartheta_{c})\sim\mathcal{N}(\mu,\sigma^{2})$ and $\operatorname{logit}(\vartheta_{t})\sim\mathcal{N}(\mu,\sigma^{2})$, with $\mu=\operatorname{logit}(0.33)$ and $\sigma=0.5$, which corresponds to a $95\%$ central interval on the response-rate scale of approximately $(0.16,0.57)$. The fitted Beta mixture has an effective sample size of around $19$ patients per arm at this rate, which is small relative to the total trial size of $3{,}800$. The prior is therefore mildly informative. This logit-normal prior is chosen to illustrate the framework's handling of a non-conjugate prior (the case that exercises the mixture-approximation step of Section~\ref{sec:23}), rather than as a clinically elicited prior for the ADRENAL re-design. The value $\sigma=0.5$ is a representative mildly informative choice for this purpose. We deliberately centre the prior at the data-generating control rate, $\mu=\operatorname{logit}(0.33)$, so that the comparison isolates the effect of the Beta-mixture approximation rather than introducing a prior-data conflict that would confound the cross-method agreement this section is designed to test.

The three methods implement this target prior differently. The proposed framework draws $50{,}000$ samples from each arm's logit-normal prior and fits a finite Beta mixture by minimising the empirical forward KL divergence with the EM algorithm of \texttt{RBesT::automixfit} \citep[\texttt{RBesT} version 1.9-0;][]{schmidli2014,weber2026}. The within-tolerance rule of Section~\ref{sec:23}, with $\varepsilon=10^{-3}$, selects $\hat{L}=2$ components for both arms. The fitted mixture has approximate weights $(0.57,0.43)$ and Beta parameters $(7.9,12.7)$ and $(8.5,21.9)$. Given that the arm-specific priors are identical, the fitted mixtures are also identical up to component labelling. We provide the full prior-approximation diagnostics for this example in~\ref{sec:A1}. \texttt{adaptr} is supplied with the target logit-normal prior directly through a custom posterior-sampling routine that draws from each arm's posterior by inverse-CDF sampling on a fine grid of $\operatorname{logit}(\vartheta)$, with no Beta-mixture intermediate, and therefore evaluates a near-exact posterior at each look. \texttt{BATSS} expresses the prior on the regression coefficients of the model $y\sim$\,arm. Writing $\beta_{0}=\operatorname{logit}(\vartheta_{c})$ and $\beta_{1}=\operatorname{logit}(\vartheta_{t})-\operatorname{logit}(\vartheta_{c})$, the independent logit-normal on the arms induces marginal distributions $\beta_{0}\sim\mathcal{N}(\mu,\sigma^{2})$ and $\beta_{1}\sim\mathcal{N}(0,2\sigma^{2})$ with cross-covariance $\operatorname{Cov}(\beta_{0},\beta_{1})=-\sigma^{2}$. As \texttt{INLA}'s \texttt{control.fixed} mechanism supports only independent normal priors on fixed effects, we drop the cross-covariance and supply the marginal-matched independent normal on $(\beta_{0},\beta_{1})$, the closest \texttt{INLA}-representable approximation to the target logit-normal on the arms. 

\begin{table}[!ht]
  \centering
  \begin{tabular}{rllrrrr}
    \toprule
    $K$ & Method & $R$ & Type I error & MC SE & $\mathbb{E}(N)$ & Time (s) \\
    \midrule
    $1$ & \texttt{BATSS} & $5{,}000$ & 0.92\,\% & 0.135\,pp & 3,800 & 2,087 \\
    $1$ & \texttt{adaptr} & $5{,}000$ & 1.10\,\% & 0.148\,pp & 3,800 & 11.9 \\
    $1$ & \texttt{adaptr} & $10^{6}$ & 0.99\,\% & 0.010\,pp & 3,800 & 1,883 \\
    $1$ & Proposed & $5{,}000$ & 0.98\,\% & 0.139\,pp & 3,800 & 3.0 \\
    $1$ & Proposed & $10^{6}$ & 0.96\,\% & 0.010\,pp & 3,800 & 8.0 \\
    \midrule
    $3$ & \texttt{BATSS} & $5{,}000$ & 2.22\,\% & 0.208\,pp & 3,436 & 5,645 \\
    $3$ & \texttt{adaptr} & $5{,}000$ & 2.10\,\% & 0.203\,pp & 3,456 & 14.7 \\
    $3$ & \texttt{adaptr} & $10^{6}$ & 2.19\,\% & 0.015\,pp & 3,455 & 2,732 \\
    $3$ & Proposed & $5{,}000$ & 2.20\,\% & 0.207\,pp & 3,458 & 6.0 \\
    $3$ & Proposed & $10^{6}$ & 2.20\,\% & 0.015\,pp & 3,455 & 11.7 \\
    \midrule
    $5$ & \texttt{BATSS} & $5{,}000$ & 2.90\,\% & 0.237\,pp & 3,231 & 8,822 \\
    $5$ & \texttt{adaptr} & $5{,}000$ & 2.68\,\% & 0.228\,pp & 3,262 & 18.0 \\
    $5$ & \texttt{adaptr} & $10^{6}$ & 2.90\,\% & 0.017\,pp & 3,264 & 3,487 \\
    $5$ & Proposed & $5{,}000$ & 2.84\,\% & 0.235\,pp & 3,277 & 5.8 \\
    $5$ & Proposed & $10^{6}$ & 2.94\,\% & 0.017\,pp & 3,260 & 19.0 \\
    \midrule
    $7$ & \texttt{BATSS} & $5{,}000$ & 3.46\,\% & 0.258\,pp & 3,118 & 12,515 \\
    $7$ & \texttt{adaptr} & $5{,}000$ & 3.24\,\% & 0.250\,pp & 3,136 & 24.8 \\
    $7$ & \texttt{adaptr} & $10^{6}$ & 3.44\,\% & 0.018\,pp & 3,138 & 4,300 \\
    $7$ & Proposed & $5{,}000$ & 3.76\,\% & 0.269\,pp & 3,150 & 9.8 \\
    $7$ & Proposed & $10^{6}$ & 3.49\,\% & 0.018\,pp & 3,132 & 20.0 \\
    \midrule
    $9$ & \texttt{BATSS} & $5{,}000$ & 4.28\,\% & 0.286\,pp & 3,007 & 15,062 \\
    $9$ & \texttt{adaptr} & $5{,}000$ & 3.72\,\% & 0.268\,pp & 3,056 & 23.8 \\
    $9$ & \texttt{adaptr} & $10^{6}$ & 3.86\,\% & 0.019\,pp & 3,037 & 5,772 \\
    $9$ & Proposed & $5{,}000$ & 3.24\,\% & 0.250\,pp & 3,037 & 14.5 \\
    $9$ & Proposed & $10^{6}$ & 3.84\,\% & 0.019\,pp & 3,040 & 36.0 \\
    \bottomrule
  \end{tabular}
  \caption{Type I error rate estimates under the logit-normal prior. As Table~\ref{tab:331}, but replacing the independent $\operatorname{Beta}(1,1)$ priors with an informative independent logit-normal prior on each arm, $\operatorname{logit}(\vartheta)\sim\mathcal{N}(\mu,\sigma^{2})$ with $\mu=\operatorname{logit}(0.33)$ and $\sigma=0.5$. The proposed framework approximates this prior by a finite Beta mixture via the prior-approximation step of Section~\ref{sec:23}. \texttt{adaptr} is supplied with the logit-normal prior directly through inverse-CDF sampling on a fine grid of $\operatorname{logit}(\vartheta)$, with no Beta-mixture intermediate. \texttt{BATSS} is supplied with the closest \texttt{INLA}-representable approximation, namely independent normal priors on the regression coefficients of the model $y\sim$\,group with marginal-matched moments $\beta_{0}\sim\mathcal{N}(\mu,\sigma^{2})$ and $\beta_{1}\sim\mathcal{N}(0,2\sigma^{2})$. The cross-covariance $\operatorname{Cov}(\beta_{0},\beta_{1})=-\sigma^{2}$ implied by the independent logit-normal on the arms cannot be expressed in \texttt{INLA}'s fixed-effects framework.}
  \label{tab:333}
\end{table}
\begin{table}[!ht]
  \centering
  \begin{tabular}{rllrrrr}
    \toprule
    $K$ & Method & $R$ & Power & MC SE & $\mathbb{E}(N)$ & Time (s) \\
    \midrule
    $1$ & \texttt{BATSS} & $5{,}000$ & 84.32\,\% & 0.514\,pp & 3,800 & 2,093 \\
    $1$ & \texttt{adaptr} & $5{,}000$ & 84.38\,\% & 0.513\,pp & 3,800 & 13.0 \\
    $1$ & \texttt{adaptr} & $10^{6}$ & 84.25\,\% & 0.036\,pp & 3,800 & 2,376 \\
    $1$ & Proposed & $5{,}000$ & 84.04\,\% & 0.518\,pp & 3,800 & 2.9 \\
    $1$ & Proposed & $10^{6}$ & 84.41\,\% & 0.036\,pp & 3,800 & 5.6 \\
    \midrule
    $3$ & \texttt{BATSS} & $5{,}000$ & 86.64\,\% & 0.481\,pp & 2,498 & 4,099 \\
    $3$ & \texttt{adaptr} & $5{,}000$ & 86.08\,\% & 0.490\,pp & 2,530 & 14.5 \\
    $3$ & \texttt{adaptr} & $10^{6}$ & 86.43\,\% & 0.034\,pp & 2,520 & 2,736 \\
    $3$ & Proposed & $5{,}000$ & 85.64\,\% & 0.496\,pp & 2,509 & 5.6 \\
    $3$ & Proposed & $10^{6}$ & 86.49\,\% & 0.034\,pp & 2,519 & 15.2 \\
    \midrule
    $5$ & \texttt{BATSS} & $5{,}000$ & 87.46\,\% & 0.468\,pp & 2,205 & 6,057 \\
    $5$ & \texttt{adaptr} & $5{,}000$ & 87.04\,\% & 0.475\,pp & 2,242 & 16.5 \\
    $5$ & \texttt{adaptr} & $10^{6}$ & 87.44\,\% & 0.033\,pp & 2,237 & 3,225 \\
    $5$ & Proposed & $5{,}000$ & 87.52\,\% & 0.467\,pp & 2,231 & 7.4 \\
    $5$ & Proposed & $10^{6}$ & 87.46\,\% & 0.033\,pp & 2,235 & 17.7 \\
    \midrule
    $7$ & \texttt{BATSS} & $5{,}000$ & 88.56\,\% & 0.450\,pp & 2,017 & 8,202 \\
    $7$ & \texttt{adaptr} & $5{,}000$ & 87.62\,\% & 0.466\,pp & 2,108 & 18.6 \\
    $7$ & \texttt{adaptr} & $10^{6}$ & 87.94\,\% & 0.033\,pp & 2,092 & 3,568 \\
    $7$ & Proposed & $5{,}000$ & 88.68\,\% & 0.448\,pp & 2,067 & 6.6 \\
    $7$ & Proposed & $10^{6}$ & 87.95\,\% & 0.033\,pp & 2,092 & 21.4 \\
    \midrule
    $9$ & \texttt{BATSS} & $5{,}000$ & 88.54\,\% & 0.450\,pp & 1,929 & 9,534 \\
    $9$ & \texttt{adaptr} & $5{,}000$ & 88.84\,\% & 0.445\,pp & 1,960 & 20.2 \\
    $9$ & \texttt{adaptr} & $10^{6}$ & 88.10\,\% & 0.032\,pp & 2,002 & 3,962 \\
    $9$ & Proposed & $5{,}000$ & 87.96\,\% & 0.460\,pp & 1,996 & 15.9 \\
    $9$ & Proposed & $10^{6}$ & 88.17\,\% & 0.032\,pp & 1,998 & 29.7 \\
    \bottomrule
  \end{tabular}
  \caption{Power estimates under the logit-normal prior. As Table~\ref{tab:333}, but reporting the estimated power rather than the type I error rate.}
  \label{tab:334}
\end{table}

The matched-budget \texttt{BATSS}, \texttt{adaptr} and proposed estimates of the type I error rate and the power agree to within a few combined Monte Carlo SEs of the two estimators being compared, and all three bracket the high-precision estimate produced by the proposed framework at $R=10^{6}$. The operating characteristics follow the same monotone trends as in Section~\ref{sec:331}: the type I error rate increases from $0.96\%$ at $K=1$ to $3.84\%$ at $K=9$ under binding futility at $\gamma^{F}=0.90$, and the expected sample size under $H_{1}$ shrinks from $3{,}800$ at $K=1$ to approximately $2{,}000$ at $K=9$. Compared to the $\operatorname{Beta}(1,1)$ baseline of Tables~\ref{tab:331} and~\ref{tab:332}, the logit-normal prior gives slightly more conservative type I error rates and modestly lower power, consistent with the prior's mild informativeness pulling posterior tails away from the decision boundary. That the three method-specific prior translations (a finite Beta mixture for the proposed framework, exact inverse-CDF sampling for \texttt{adaptr}, and a marginal-matched independent normal on the logit-scale coefficients for \texttt{BATSS}) all reproduce these operating characteristics within Monte Carlo error provides empirical reassurance that the finite-mixture approximation introduces no systematic bias at the decision-rule resolution. Since the logit-normal posterior is non-conjugate, here the proposed framework's Beta mixture uses a genuine two-component approximation ($\hat{L}=2$) rather than collapsing to a single exact posterior, so the head-to-head against \texttt{adaptr}'s exact inverse-CDF posterior at the high-precision $R=10^{6}$ budget is a direct test of the approximation. At that budget, where each estimator's binomial SE is about $0.02$ percentage points, the two agree to within $0.05$ percentage points on the type I error rate across all five designs and to within $0.07$ percentage points on power for every design with early stopping, bounding the error introduced by the finite-mixture approximation itself, as distinct from Monte Carlo noise, below one tenth of a percentage point on the decision probabilities.

The proposed framework remains roughly three orders of magnitude faster than \texttt{BATSS} under the logit-normal prior (approximately $600\times$ to $1{,}500\times$ across the GSDs, Tables~\ref{tab:333} and~\ref{tab:334}, last column), although the per-design speedup is modestly smaller than under the conjugate prior ($3{,}700\times$ to $6{,}600\times$ in Section~\ref{sec:331}), reflecting the slightly higher per-trial cost of the mixture-weighted posterior tail probability under multiple components. The prior-approximation step itself is a one-off pre-processing cost (a few seconds on $50{,}000$ samples) that does not scale with $R$, so it contributes a constant additive overhead that is dwarfed by the simulation pass at any practical budget. The precomputation advantage of Section~\ref{sec:24} is unchanged. Once the per-arm Beta mixture is fitted, the cached per-look tail probabilities are still indexed by the simulated event counts and the candidate effect thresholds alone, and the joint-calibration discussion of Section~\ref{sec:34} applies verbatim under the non-conjugate prior.

\subsection{Operating characteristics across the design grid}
\label{sec:34}
\begin{figure}[!ht]
  \centering
  \includegraphics[width=\linewidth]{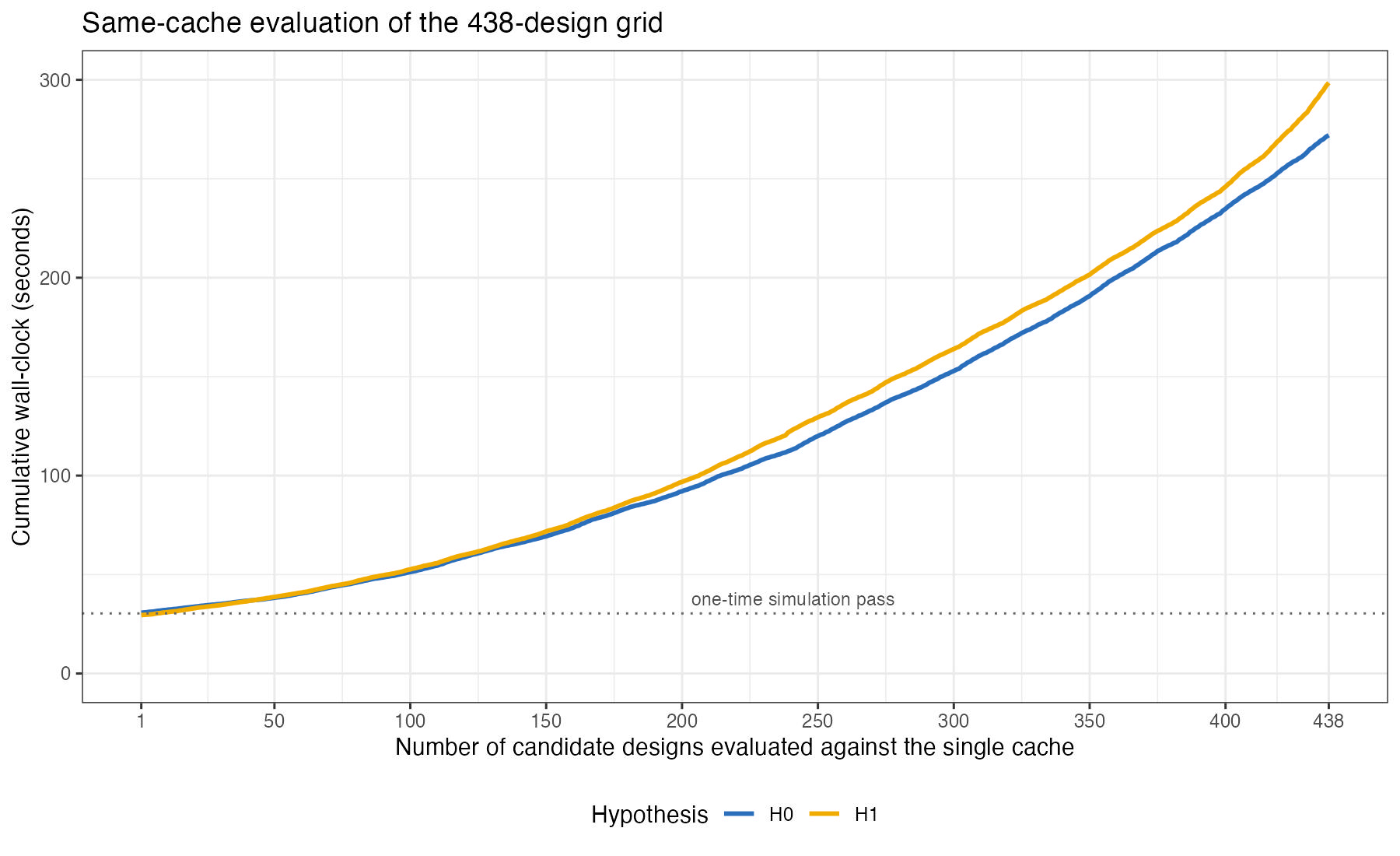}
  \caption{Same-cache evaluation of the 438-design grid. A single Monte Carlo simulation pass is run under each hypothesis at the union of all candidate look times (the one-time simulation cost, shown as the dotted intercept at about $30$ seconds). The 438 designs are evaluated one at a time, and the cumulative wall-clock time is recorded after each. Each additional design adds about $0.55$ seconds under $H_{0}$ and $0.61$ seconds under $H_{1}$, so the full grid is calibrated against the one cache in approximately four minutes per hypothesis, with no further simulation. By contrast, \texttt{BATSS} and \texttt{adaptr} treat the design skeleton and the decision thresholds as simulator inputs and would re-simulate all 438 designs from scratch. Measured on a quad-core workstation (8 logical threads) at $R=10^{6}$. This panel is a separate illustrative run whose absolute timings differ slightly from the full-grid benchmark reported in Section~\ref{sec:331}, while exhibiting the same linear-in-design scaling.}
  \label{fig:341}
\end{figure}

To exercise the framework across its full range, we evaluate the full 438-design grid under both $H_{0}$ and $H_{1}$, spanning $K\in\{1,2,\ldots,10\}$, the efficacy thresholds $\gamma^{E} \in \{0.90,0.95,0.99\}$ and futility thresholds $\gamma^{F} \in \{0.85,0.90,0.95,1.00\}$, the single- and dual-criterion efficacy rules, and both binding and non-binding futility, with the within-$K$ look spacing held equal throughout. For each hypothesis, all designs are evaluated against a single cached Monte Carlo pass, so the full grid is swept in minutes rather than re-simulated design by design (Figure~\ref{fig:341}), the operational illustration of the cache invariance of Lemma~\ref{lem:241}. 

The resulting type I error, type II error, and expected-sample-size surfaces are reported in full in Figures~\href{run:./ZH2023_Supplemental_Material.pdf}{S1}--\href{run:./ZH2023_Supplemental_Material.pdf}{S5}. These surfaces illustrate the expected patterns: under constant thresholds, the type I error rate increases with the number of looks (Figures~\href{run:./ZH2023_Supplemental_Material.pdf}{S1} and~\href{run:./ZH2023_Supplemental_Material.pdf}{S2} under binding and non-binding futility), and the stringent threshold $\gamma^{E}=0.99$ that controls the error rate at $K=1$ loses that control as $K$ increases. This climb reflects the multiplicity of repeated testing against a fixed posterior-probability boundary, as discussed in Section~\ref{sec:4}. The dual-criterion efficacy rule of \citet{gsponer2014} is uniformly more conservative than the single-criterion rule, yielding lower type I error and higher expected sample size. Type II error (Figure~\href{run:./ZH2023_Supplemental_Material.pdf}{S3}) and the expected sample size under $H_{0}$ and $H_{1}$ (Figures~\href{run:./ZH2023_Supplemental_Material.pdf}{S4} and~\href{run:./ZH2023_Supplemental_Material.pdf}{S5}) both fall as looks are added, with diminishing returns. The same cached pass also provides the per-look probabilities of stopping for efficacy or futility at no additional cost. We report this per-look breakdown, together with the cumulative expected sample size, for the five-look re-design with $\gamma^{E}=0.99$ and $\gamma^{F}=0.90$ in Section~\href{run:./ZH2023_Supplemental_Material.pdf}{S5} as a worked numerical illustration. 

\subsection{Threshold calibration to the design targets}
\label{sec:35}
\begin{figure}[!ht]
  \centering
  \includegraphics[width=\linewidth]{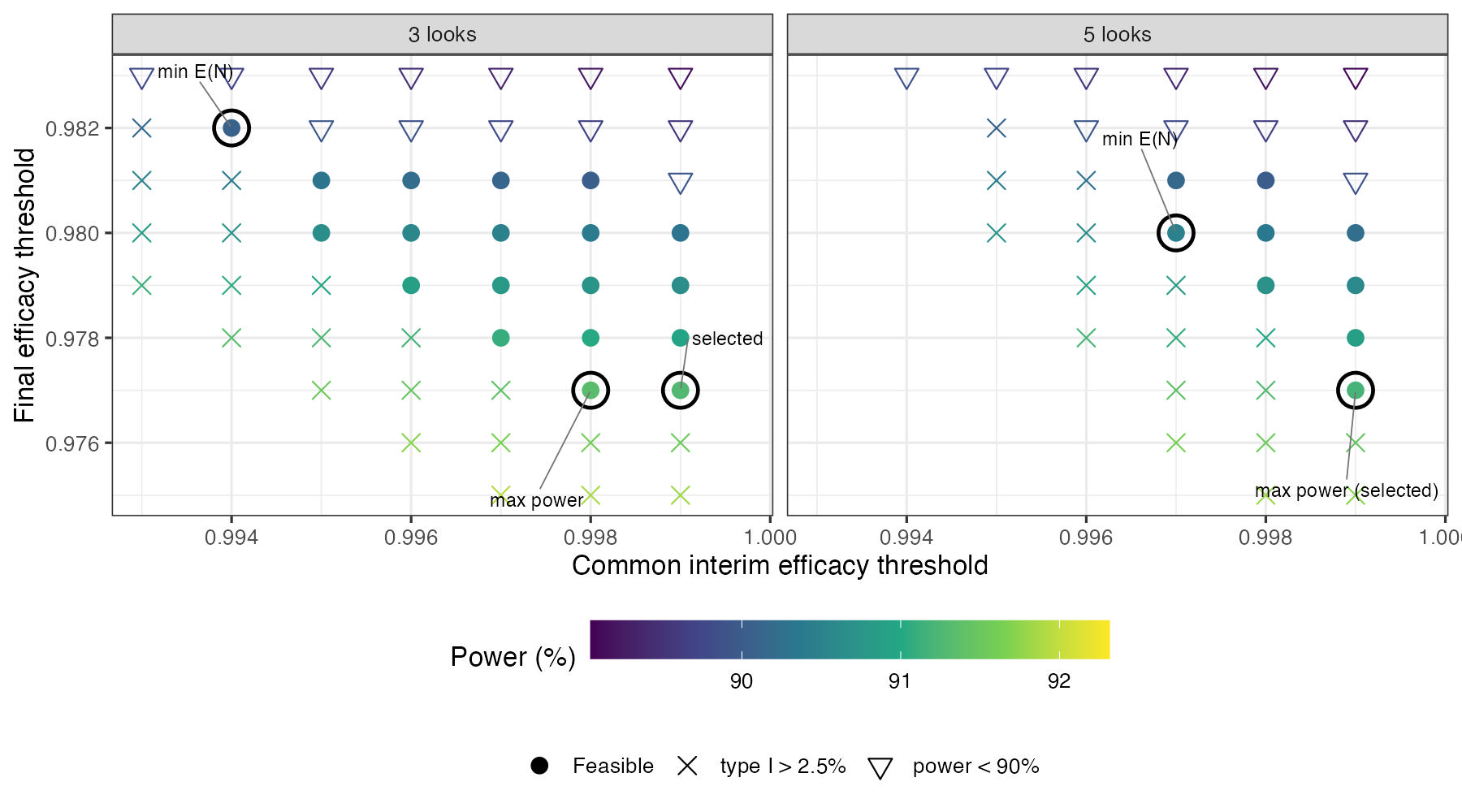}
  \caption{Feasible single-criterion Haybittle--Peto efficacy-threshold pairs for the ADRENAL re-design with $\gamma^{F}=0.90$, evaluated from the precomputed cache, for the three-look (left) and five-look (right) designs. Each point is a common interim efficacy threshold (horizontal axis) paired with a final efficacy threshold (vertical axis), coloured by power. A design is feasible (filled point) when its non-binding type I error supremum over the plausible control-rate band $\vartheta \in \{0.15,0.20,0.25,0.30,0.33,0.36,0.40,0.45,0.50\}$ is at most $2.5\%$ and its power at the design alternative is at least $90\%$. Crosses denote a band supremum above $2.5\%$ and open triangles denote power below $90\%$, so the feasible family is bounded by the $2.5\%$ type I and $90\%$ power contours. Greater interim stringency permits a lower final threshold, trading power against $\mathbb{E}(N\mid H_{1})$. Circles mark three reference designs within the feasible family: the maximum-power design, the minimum-$\mathbb{E}(N\mid H_{1})$ design, and the selected design, which takes the most stringent interim threshold $0.999$ for the largest robustness margin. At five looks the maximum-power and selected designs coincide.}
  \label{fig:351}
\end{figure}

Across the fixed-threshold grid in Section~\ref{sec:34}, no configuration simultaneously satisfies the original ADRENAL targets of $\alpha\leq 2.5\%$ (one-sided) and $1-\beta\geq 90\%$. We therefore calibrate the single-criterion efficacy thresholds on a Haybittle--Peto schedule, sweeping a common stringent interim threshold $\gamma^{E}_{\mathrm{int}}=\gamma_{1}^{E}=\cdots=\gamma_{K-1}^{E}$ and a separate near-nominal final threshold $\gamma^{E}_{K}$ on a grid refined towards one at resolution $0.001$. A single precomputed simulation pass evaluates the full grid from the cache in a fraction of a second, yielding the full operating-characteristic surface at negligible additional cost. Imposing the type I error and power constraints identifies a feasible family rather than a unique design. For the three- and five-look schedules, Figure~\ref{fig:351} shows the admissible band of $(\gamma^{E}_{\mathrm{int}},\gamma^{E}_{K})$ pairs, bounded by the $\alpha \leq 2.5\%$ and $1-\beta \geq 90\%$ contours. Greater interim stringency preserves more type I error for the final analysis and therefore permits a lower final threshold. Power is largely insensitive to the common futility threshold over $[0.85,1]$, so we fix $\gamma^{F}=0.90$, corresponding to futility when the posterior probability of benefit falls below $10\%$.

Calibration amounts to picking a point from the feasible family based on a chosen objective. Because the cache provides the operating characteristics of every admissible design, the objective may target power, expected sample size or study duration under the null or alternative, or a composite criterion. Here we show the step with maximum power.  
We additionally control the non-binding type I error not only at the design null but also as its supremum over the plausible control-rate band $\vartheta \in \{0.15,0.20,0.25,0.30,0.33,0.36,0.40,0.45,0.50\}$. Each control rate needs a separate simulation pass, but that pass evaluates every threshold pair, so the band supremum requires one simulation per control rate rather than one per design. The family's most lenient designs achieve the greatest power but exceed the type I error target over the band. Among the band-controlled designs, whose power is nearly constant, we choose the most stringent interim threshold, $0.999$, which provides the largest robustness margin. Minimising the expected sample size would instead favour the opposite corner of the family, with a less stringent interim and a more stringent final. As selecting a design near the boundary from several thousand cached candidates may introduce winner's-curse bias, we verify each selected design independently on its own analysis schedule.

The calibrated three-look and five-look designs adopt $\boldsymbol{\gamma}_{1:3}^{E}=(0.999,0.999,0.977)$ and $\boldsymbol{\gamma}_{1:5}^{E}=(0.999,0.999,0.999,0.999,0.977)$, respectively, with $\gamma^{F}=0.90$. Evaluated on their own analysis schedules, both achieve approximately $91\%$ power while controlling the type I error within $2.5\%$ under binding and non-binding futility. The five-look design has a smaller expected sample size under the alternative ($\mathbb{E}(N\mid H_{1})=2{,}972$ versus $3{,}164$), reflecting its additional early-stopping opportunities. The corresponding values under the null are $3{,}295$ and $3{,}474$. We report these designs to illustrate the calibration procedure rather than represent recommended optima, since the ADRENAL re-design is used primarily to exercise the method. The same cache calibrates a design at every $K\in\{1,2,\ldots,10\}$. The resulting frontier, reported in Section~\href{run:./ZH2023_Supplemental_Material.pdf}{S6}, maintains power near $91\%$ while expected sample size decreases monotonically with the number of looks. When study duration is of interest, under a fixed accrual rate $\lambda$, the expected sample size follows from Eq.~(\ref{eqn:246}) as $\mathbb{E}(T)=\mathbb{E}(N)/\lambda+90$ days, including the 90-day follow-up for the ADRENAL endpoint.

We assess robustness beyond the calibration point in two ways, each using $R=10^{6}$. First, holding $\delta=0$ and varying $\vartheta \in \{0.15,0.20,0.25,0.30,0.33,0.36,0.40,0.45,0.50\}$, the type I error rate remains below $2.5\%$ under both futility conventions. The non-binding suprema are $2.41\%$ and $2.46\%$ for the three- and five-look designs, respectively, with corresponding binding suprema $2.40\%$ and $2.45\%$. With $\vartheta=0.33$, the power increases monotonically with the effect size and reaches approximately $91\%$ at the design alternative $\delta=-0.05$. Second, across the canonical seed (\texttt{SEED}=21) and nine additional independent seeds, both designs maintain the type I error rate below $2.5\%$ and the power above $90\%$ under both futility conventions, with seed-to-seed variability matching the binomial Monte Carlo SE. Because these guarantees are empirical rather than uniform over the null, such plausibility sweeps should accompany any submission, and the precomputation strategy makes them inexpensive.

\subsection{Monte Carlo precision and simulation budget}
\label{sec:36}
\begin{figure}[!ht]
	\centering
	\includegraphics[width=\linewidth]{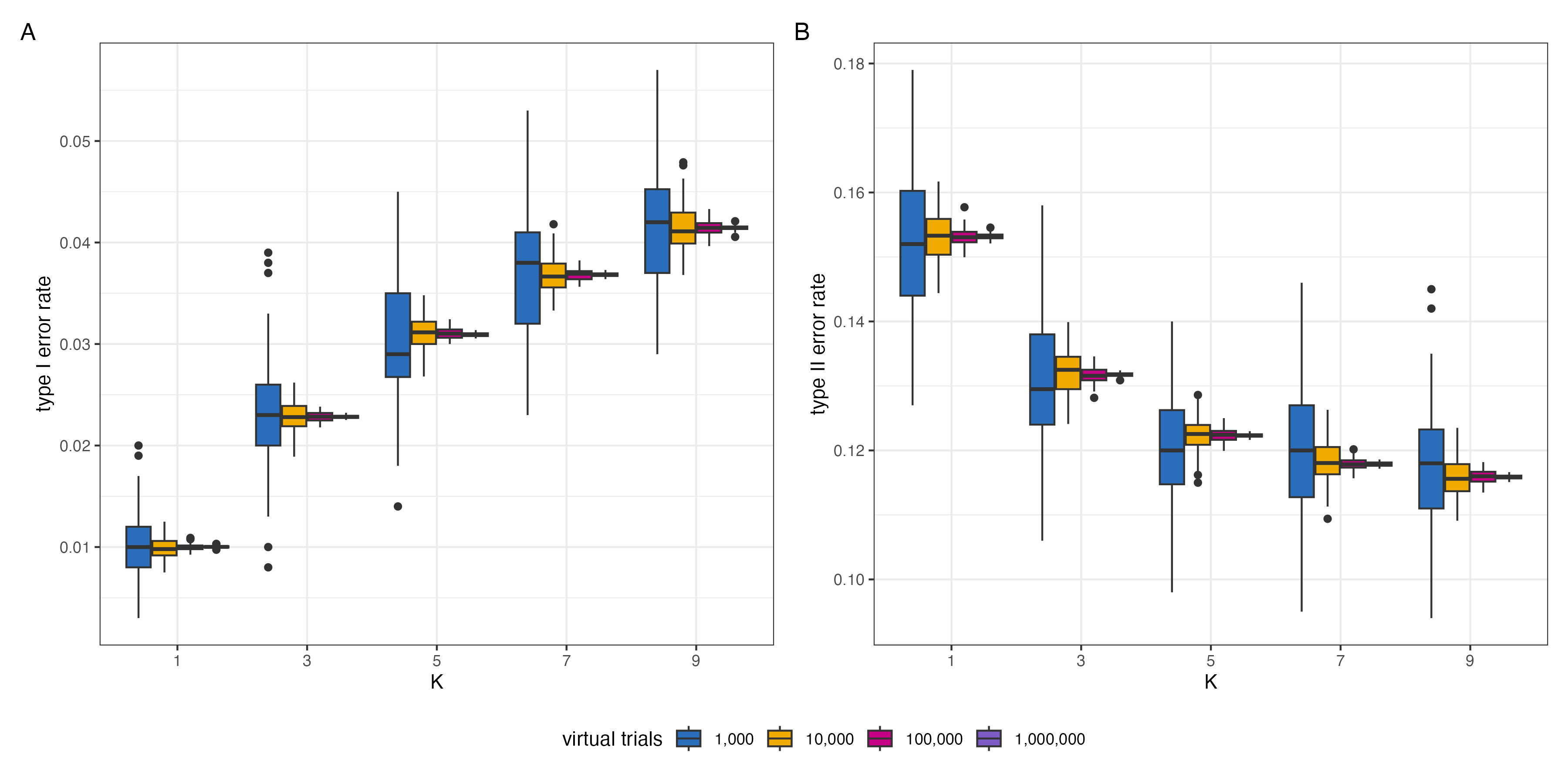}
	\caption{Boxplots showing the variability in estimated type I and type II error rates for Bayesian GSDs with $K \in \{1,3,5,7,9\}$ and varying virtual-trial budgets ($10^{3}$, $10^{4}$, $10^{5}$ and $10^{6}$), under binding futility. The decision rule stops the trial early for efficacy when the posterior probability that hydrocortisone reduces 90-day all-cause mortality exceeds 99\%, and for binding futility when this posterior probability falls below 10\%. The prior is $\operatorname{Beta}(1,1)$ in both arms. Each scenario is replicated 100 times, using disjoint (non-overlapping) random-number streams across replicates and across virtual-trial budgets, to quantify Monte Carlo variability. Panel A: type I error rate, with virtual trials drawn at the $H_{0}$ data-generating values $(\vartheta,\delta)=(0.33,0)$. Panel B: type II error rate, with virtual trials drawn at the $H_{1}$ data-generating values $(\vartheta,\delta)=(0.33,-0.05)$.}
	\label{fig:361}
\end{figure}

To explore how simulation size influences the precision of operating characteristics estimates, we examine Bayesian GSDs for the ADRENAL trial with the following decision rules. The trial stops early for efficacy when $\mathbb{P}(\Delta<0\mid\,\cdot\,)>0.99$, that is, when the posterior probability that hydrocortisone reduces 90-day all-cause mortality exceeds $99\%$. It stops early for futility when $\mathbb{P}(\Delta>0\mid\,\cdot\,)>0.90$. 
We evaluate the overall type I and type II error rates for designs with $K\in\{1,3,5,7,9\}$ total analyses, namely a fixed-sample design at $K=1$ and GSDs with two, four, six and eight equally spaced interim looks for $K\in\{3,5,7,9\}$. Each scenario is simulated using $10^{3}$, $10^{4}$, $10^{5}$, and $10^{6}$ virtual trials per replicate, with 100 independent replicates specified to quantify Monte Carlo variability. Boxplots of the resulting distributions of estimated type I and type II error rates are shown in Figure~\ref{fig:361}, with the corresponding numerical summaries reported in Table~\href{run:./ZH2023_Supplemental_Material.pdf}{S4}. Since each design skeleton here is re-simulated with its own independent random-number streams, the mean type I error rate for a given $K$ differs by up to about $0.02$ percentage points from the single-seed benchmark entry of Table~\ref{tab:331} (for instance, $3.09\%$ against $3.11\%$ at $K=5$ and $R=10^{6}$), a difference that lies within the binomial Monte Carlo SE at this budget.

Figure~\ref{fig:361}A demonstrates that the variability of the type I error rate estimates increases with the number of interim looks for a fixed number of virtual trials. At $R=10^{3}$, the Monte Carlo SE roughly doubles from $0.31$ percentage points at $K=1$ to $0.64$ percentage points at $K=9$, mirroring the corresponding climb in the mean type I error rate from $1.02\%$ to $4.13\%$ under uncalibrated decision thresholds. Increasing the number of virtual trials to $10^{4}$, $10^{5}$, and then $10^{6}$ shrinks the Monte Carlo SE by a factor of approximately $3.16$ at each step, as expected for a binary estimator, so that the largest SE at $R=10^{5}$ is $0.07$ percentage points (at $K=9$) and at $R=10^{6}$ it is $0.02$ percentage points (at $K=9$).

In contrast, Figure~\ref{fig:361}B shows that the variability of the type II error rate estimates is almost flat across the number of interim looks. The Monte Carlo SE ranges from $0.91$ to $1.11$ percentage points at $R=10^{3}$ for every $K\in\{1,3,5,7,9\}$, shrinks to between $0.09$ and $0.12$ percentage points at $R=10^{5}$, and further to between $0.03$ and $0.04$ percentage points at $R=10^{6}$. The SE at $R=10^{3}$ is essentially unchanged across $K$ (up to sampling noise). This pattern demonstrates the fact that the estimated type II error rates are broadly similar across designs (ranging $11.6\%$ at $K=9$ to $15.3\%$ at $K=1$) and therefore have comparable binomial variance.

Two practical implications follow. First, at a fixed simulation budget, the Monte Carlo SE of the type I error estimate is more sensitive to the number of interim looks than that of the type II error estimate. Hence, the simulation budget should be chosen with the largest number of analyses in the candidate design grid in mind. Second, $R=10^{5}$ already reduces the Monte Carlo SE of the type I error estimate below $0.1$ percentage points across all design skeletons we consider, providing sufficient precision for reliable threshold calibration against a one-sided target $\alpha=2.5\%$. Increasing to $R=10^{6}$ further reduces this SE to below $0.02$ percentage points and, given the low computational burden of the framework, provides a practical default with additional margin for calibration supporting regulatory review, where operating characteristics are typically required across a plausibility set of data-generating values (Section~\ref{sec:35}).

To illustrate the consequences of an underpowered simulation budget, we gate the five design skeletons against the $\alpha\leq 2.5\%$ target, using the corresponding $R=10^{6}$ estimate as the reference value and exploring how often the smaller-budget replicates fall on the wrong side ($\alpha>2.5\%$). The design closest to the gate, $K=3$, has a reference type I error rate of $2.28\%$, $0.22$ percentage points below the target. It is misclassified in $27\%$ of the $100$ replicates at $R=10^{3}$ and $9\%$ at $R=10^{4}$, but in none at $R=10^{5}$ or above. Designs farther from the gate are misclassified less often ($18\%$ at $K=5$ and $3\%$ at $K=7$, both at $R=10^{3}$, and never for $K=1$ or $K=9$ at any budget), and none at all beyond $R=10^{4}$. Under-budgeted calibration thus risks both false acceptance and false rejection of designs whose true type I error rate lies within approximately one Monte Carlo SE of the target, whereas $R=10^{5}$ eliminates this risk for the design skeletons considered here.


\section{Discussion}
\label{sec:4}
In this work, we have introduced a semi-simulation framework for the rapid evaluation and calibration of Bayesian GSDs. The contribution has two complementary components. The first is a closed-form per-look posterior update that works for any user-specified prior admitting an accurate finite-mixture-of-conjugates approximation. The second is a precomputation strategy that decouples the cost of simulation from any particular design evaluation, so a single Monte Carlo pass at the union of all candidate look times supports the calibration over a large grid of candidate designs at marginal computational cost.

The first component, closed-form per-look updating through a finite-mixture-of-conjugates prior approximation, removes the need for MCMC within the trial-simulation loop while maintaining flexibility in prior specification. In the ADRENAL trial re-design under both a uniform conjugate $\operatorname{Beta}(1,1)$ prior and an informative non-conjugate logit-normal prior, the proposed framework reproduces the operating characteristics of \texttt{BATSS} and \texttt{adaptr} within Monte Carlo error across designs with up to nine analyses. At a matched high-precision budget of $R=10^{6}$, where Monte Carlo noise is negligible, the proposed framework and \texttt{adaptr} agree on the estimated type I error rate to within $0.05$ percentage points across all designs under both priors. At matched simulation budgets, it is around three orders of magnitude faster than \texttt{BATSS} per GSD (roughly $3{,}700\times$ to $6{,}600\times$ under the conjugate prior and $600\times$ to $1{,}500\times$ under the logit-normal prior) and, under the conjugate prior, $7\times$ to $16\times$ faster than \texttt{adaptr}, with the advantage widening to several hundredfold at the $R=10^{6}$ calibration budget. The simulation-budget study additionally shows that $10^{5}$ simulated trials already provide Monte Carlo precision adequate for reliable threshold calibration, while the low computational cost of the framework makes $10^{6}$ simulated trials a practical routine default for more precise calibration.

The second component, the precomputation strategy, distinguishes this work from existing methods and is where the largest practical gain arises. In design-stage applications, the objective is not to evaluate a single fixed Bayesian GSD, but to search over candidate numbers and timings of interim analyses, efficacy and futility thresholds, and binding or non-binding futility rules under prespecified type I and type II error targets. Existing tools treat these choices as simulator inputs, so the wall-clock cost of calibration scales roughly linearly in the number of candidate designs evaluated. Our strategy avoids this repeated simulation cost by exploiting the fact that, for a fixed prior and analysis model, the per-look posterior tail probability $\mathbb{P}(\Delta>\delta \mid \,\cdot\,)$ relies only on the simulated event counts and the candidate effect threshold $\delta$.
A single simulation pass at the union of all candidate look times is therefore sufficient, after which each design is evaluated by sweeping its stopping rules over the cached posterior probabilities, so that joint calibration of decision thresholds and design skeleton reduces to a sweep over the candidate grid. This is related to the common-random-numbers principle \citep{law2015}, but the conjugate structure makes the reuse exact and threshold-free. A design that changes only its look subset, its posterior-probability thresholds or its futility convention introduces no re-estimation and no additional Monte Carlo noise. In the ADRENAL example, the full 438-design grid is evaluated in about eight minutes on eight logical CPUs, and the calibrated three- and five-look designs meet the original targets of $\alpha\leq 2.5\%$ and $1-\beta\geq 90\%$. The precomputation strategy makes routine joint calibration of Bayesian GSDs against prespecified error targets accessible on commodity hardware.

The results also clarify why calibration is necessary. For a fixed design and data-generating distribution, the type I error rate is a deterministic quantity, and the simulation budget controls only the precision with which it is estimated. Under uncalibrated thresholds, the type I error rate increases from around $1.0\%$ for a single-look design ($K=1$) to $4.1\%$ for a nine-look design ($K=9$) at $\gamma^{E}=0.99$ and $\gamma^{F}=0.90$. This increase across designs is the familiar multiplicity of repeated significance testing \citep{jennison1999}, in that applying the same posterior-probability threshold at every analysis tests the null repeatedly, so the cumulative probability of crossing it under $H_{0}$ grows with the total number of analyses. A constant posterior-probability boundary has the same pattern as a Pocock-type boundary \citep{pocock1977}, reflecting the well-documented correspondence between fixed Bayesian posterior-probability boundaries and frequentist error-spending boundaries \citep{lewis1994,spiegelhalter2004}. The distinction is calibration, in that a Pocock spending function adjusts that constant boundary to the number of analyses so that the total type I error remains at the nominal level for every $K$, whereas a fixed, uncalibrated threshold does not and therefore inflates as looks are added. This inflation is not a Monte Carlo artefact but a property of the stopping rule. The proposed framework makes such behaviour straightforward to diagnose and allows the decision thresholds to be recalibrated against prespecified type I and type II error targets. 

The proposed framework should be viewed as complementary to existing software rather than as a replacement for all current methods. As reviewed in Section~\ref{sec:1}, \texttt{gsbDesign} \citep{gerber2016}, \texttt{adaptr} \citep{granholm2022} and \texttt{BATSS} \citep{couturier2024} each occupy a specific niche: normal endpoints with normal priors, response-adaptive randomisation with fixed conjugate families, and broader priors and endpoints via \texttt{INLA}, respectively. The proposed framework instead targets fixed-allocation Bayesian GSDs with posterior-probability decision rules, where the main practical challenge is to make large-budget, reproducible operating-characteristic evaluation and calibration feasible across large design grids. This enables more precise estimation of type I error, power, and expected sample size while allowing many threshold and interim-schedule configurations to be explored. The benchmark of Section~\ref{sec:33} should therefore be interpreted as a like-for-like comparison of wall-clock cost on this shared task rather than a claim of universal superiority. The speed advantage is achieved by exploiting the specific structure of the problem: under a finite-mixture-of-conjugates prior, the posterior-probability decision rules can be evaluated by closed-form conjugate updating and low-dimensional Gauss--Legendre quadrature, avoiding repeated MCMC or approximate posterior inference within the trial-simulation loop. Beyond this per-scenario speed-up, the precomputation strategy makes the gain architectural as well, so the framework is most valuable when the objective is to jointly calibrate the design skeleton and decision thresholds against common error targets.

In practice, we recommend the following workflow. First, simulate once at the union of all candidate look times, using $R=10^{6}$ virtual trials as a routine default ($R=10^{5}$ is the smallest budget we found adequate for calibration against $\alpha\leq 2.5\%$, see Section~\ref{sec:36}), and cache the per-look posterior tail probabilities. Second, for a non-conjugate prior, fit the Beta-mixture approximation with $L_{\max}=5$ candidate components and the within-tolerance rule with $\varepsilon=10^{-3}$ of Section~\ref{sec:23}, tightening $\varepsilon$ for heavy-tailed or sharply peaked target priors. Third, sweep the cache over the candidate effect thresholds, posterior-probability thresholds, look schedules and binding or non-binding futility conventions (all of which leave the cache valid by Lemma~\ref{lem:241}), and select the design that meets the target type I error rate while optimising power, expected sample size, or expected study duration (when recruitment, follow-up, and dropout are modelled). Fourth, for a regulatory submission, report the prior-approximation diagnostics of~\ref{sec:A1} (the KL-by-$L$, density overlay, prior-tail and operating-characteristic-sensitivity checks of Figure~\ref{fig:A11} and Tables~\ref{tab:A11}--\ref{tab:A12}) and the quadrature-convergence check of~\ref{sec:A2} (repeated at the design's own sample sizes and event rates when these depart materially from the worked example) alongside the calibrated design and its operating characteristics under both futility conventions, together with the type I error rate swept over a plausibility set of nuisance-parameter values to confirm control across the null rather than at the single calibration point alone.

The ADRENAL example re-designs an existing trial, so the maximum sample size $N=3{,}800$ is given. For a new trial, however, the maximum sample size can be searched on the same cache rather than through repeated simulation. From Lemma~\ref{lem:241}, the cached per-look tail probability depends only on the data accumulated by that sample size, not on the planned maximum. Thus, any candidate design whose analysis times lie within the simulated union, including designs with smaller maxima, can be evaluated directly from a single cache. In practice, one simulation pass is run to the largest candidate maximum sample size over a union of analysis times defined in sample sizes or information fractions. The maximum sample size is then optimised jointly with the look schedule and decision thresholds, selecting the smallest maximum, or expected sample size, that meets the type I error and power constraints. A fresh simulation pass is required only for candidate maxima beyond the simulated union, so the upper bound should be set generously. 

These advantages notwithstanding, our framework has several limitations. The most fundamental is that calibration is still simulation-based and therefore subject to Monte Carlo error. At least $10^{5}$ trials per scenario are required for reliable estimation of the type I error rate \citep{fda2010}, and the resulting cost scales unfavourably with design complexity and with the size of the threshold grid \citep{broglio2022}. The stochasticity also makes principled optimisation of design parameters such as the timing of interim looks harder than in the analogous frequentist problem, where deterministic spending functions allow direct algorithmic optimisation \citep{he2025}. 

The worked examples here focus on a binary endpoint with a Beta-mixture prior, where conjugacy is most direct. Sections~\ref{sec:22} and~\ref{sec:23} extend the framework to the risk-ratio and odds-ratio scales, 
and Section~\href{run:./ZH2023_Supplemental_Material.pdf}{S1} adapts it to continuous, count and time-to-event endpoints through the normal--normal or normal--inverse-gamma, gamma--Poisson and gamma--exponential conjugate pairs, respectively, but the numerical-integration cost in those settings has not yet been benchmarked. The current implementation also does not support predictive-probability monitoring, response-adaptive randomisation or covariate adjustment. 

A further limitation is that the calibrated thresholds control the frequentist type I error rate only over the nuisance-parameter range against which they are checked. As shown in Section~\ref{sec:35}, the type I error rate can vary non-monotonically with the control event rate and can peak away from the assumed design value. Therefore, the thresholds carry no uniform guarantee over the whole null and should be verified through a plausibility sweep for the target population. The prior approximation introduces additional error, whose effect on the calibrated thresholds has not yet been quantified. \ref{sec:A1} thus recommends reporting four diagnostics alongside the operating characteristics in non-conjugate settings: the KL divergence by $L$, a density overlay of the target prior and the fitted mixture, prior tail probabilities at the decision thresholds, and the sensitivity of these summaries to $L$.

These limitations motivate several directions for future work. The most consequential is to develop a simulation-free or semi-analytical alternative for the calibration step itself, addressing the transparency, reproducibility and prespecification considerations emphasised in the draft guidance of \citet{fda2026} on Bayesian methodology. The framework could be extended to support predictive-probability monitoring and common adaptations such as response-adaptive randomisation and sample size re-estimation, and to incorporate covariate adjustment, all of which are presently outside its scope. Finally, a user-facing \texttt{R} package and a Shiny interface would further facilitate routine application by trial statisticians.

In summary, by combining a conjugate-mixture prior approximation with a precomputation strategy that decouples trial simulation from design evaluation, the proposed framework makes joint calibration of the design skeleton and decision thresholds feasible on commodity hardware. This reduces the computational barrier to implementing Bayesian GSD in confirmatory trials, while recognising that regulatory and prior-specification considerations remain relevant to their broader adoption.

\subsection*{Data availability statement}
This study uses only synthetic data generated by Monte Carlo simulation. No human-subject data are used. All R code, simulation drivers, cached simulation outputs and figure-rendering scripts required to reproduce the manuscript and its supplement are available from \url{https://github.com/zhangyi-he/BayesGSD/tree/main/SemiSimul/PostPr}, version-tagged as released with this paper. The published trial parameters of the ADRENAL trial used to motivate the re-design are taken from \citet{venkatesh2018}.

\subsection*{Funding}
The authors declare no specific funding associated with the work presented in this article.

\subsection*{Conflict of interest}
The authors declare no potential conflict of interest.

\subsection*{Ethics statement}
This work uses only simulated data and the previously published summary statistics of the ADRENAL trial. No new human-subject data were collected and no ethical approval was therefore required.



\bibliography{ZH2023_Manuscript}

@article{jennison2005,
  author={Jennison, C and Turnbull, B W},
  title={{Meta-analyses and adaptive group sequential designs in the clinical development process}},
  journal={Journal of Biopharmaceutical Statistics},
  volume={15},
  number={4},
  pages={537--558},
  year={2005}
}

@article{pallmann2018,
  author={Pallmann, P and Bedding, A W and Choodari-Oskooei, B and Dimairo, M and Flight, L and others},
  title={{Adaptive designs in clinical trials: why use them, and how to run and report them}},
  journal={BMC Medicine},
  volume={16},
  pages={29},
  year={2018},
}

@article{stevely2015,
  author={Stevely, A and Dimairo, M and Todd, S and Julious, S A and Nicholl, J and others},
  title={{An investigation of the shortcomings of the CONSORT 2010 statement for the reporting of group sequential randomised controlled trials: a methodological systematic review}},
  journal={PLoS One},
  volume={10},
  number={11},
  pages={e0141104},
  year={2015}
}

@article{judge2021,
  author={Judge, C and Murphy, R and Reddin, C and Cormican, S and Smyth, A and others},
  title={{Trends in adaptive design methods in dialysis clinical trials: a systematic review}},
  journal={Kidney Medicine},
  volume={3},
  number={6},
  pages={925--941},
  year={2021}
}

@book{jennison1999,
  author={Jennison, C and Turnbull, B W},
  title={{Group Sequential Methods with Applications to Clinical Trials}},
  publisher={Chapman \& Hall/CRC},
  address={Boca Raton},
  year={1999}
}

@book{wassmer2016,
  author={Wassmer, G and Brannath, W},
  title={{Group Sequential and Confirmatory Adaptive Designs in Clinical Trials}},
  publisher={Springer},
  address={Heidelberg},
  year={2016}
}

@article{keh2016,
  author={Keh, D and Trips, E and Marx, G and Wirtz, S P and Abduljawwad, E and others},
  title={{Effect of hydrocortisone on development of shock among patients with severe sepsis: the HYPRESS randomized clinical trial}},
  journal={JAMA},
  volume={316},
  number={17},
  pages={1775--1785},
  year={2016}
}

@article{combes2018,
  author={Combes, A and Hajage, D and Capellier, G and Demoule, A and Lavou{\'e}, S and others},
  title={{Extracorporeal membrane oxygenation for severe acute respiratory distress syndrome}},
  journal={New England Journal of Medicine},
  volume={378},
  number={21},
  pages={1965--1975},
  year={2018}
}

@article{perkins2018,
  author={Perkins, G D and Ji, C and Deakin, C D and Quinn, T and Nolan, J P and others},
  title={{A randomized trial of epinephrine in out-of-hospital cardiac arrest}},
  journal={New England Journal of Medicine},
  volume={379},
  number={8},
  pages={711--721},
  year={2018}
}

@article{haybittle1971,
  author={Haybittle, J L},
  title={{Repeated assessment of results in clinical trials of cancer treatment}},
  journal={The British Journal of Radiology},
  volume={44},
  number={526},
  pages={793--797},
  year={1971}
}

@article{peto1976,
  author={Peto, R and Pike, M C and Armitage, P and Breslow, N E and Cox, D R and others},
  title={{Design and analysis of randomized clinical trials requiring prolonged observation of each patient. I. Introduction and design}},
  journal={British Journal of Cancer},
  volume={34},
  number={6},
  pages={585--612},
  year={1976}
}

@article{pocock1977,
  author={Pocock, S J},
  title={{Group sequential methods in the design and analysis of clinical trials}},
  journal={Biometrika},
  volume={64},
  number={2},
  pages={191--199},
  year={1977}
}

@article{obrien1979,
  author={O'Brien, P C and Fleming, T R},
  title={{A multiple testing procedure for clinical trials}},
  journal={Biometrics},
  volume={35},
  number={3},
  pages={549--556},
  year={1979}
}

@article{lan1983,
  author={Lan, K K G and DeMets, D L},
  title={{Discrete sequential boundaries for clinical trials}},
  journal={Biometrika},
  volume={70},
  number={3},
  pages={659--663},
  year={1983}
}

@article{kim1987,
  author={Kim, K and DeMets, D L},
  title={{Design and analysis of group sequential tests based on the type I error spending rate function}},
  journal={Biometrika},
  volume={74},
  number={1},
  pages={149--154},
  year={1987}
}

@article{hwang1990,
  author={Hwang, I K and Shih, W J and De Cani, John S},
  title={{Group sequential designs using a family of type I error probability spending functions}},
  journal={Statistics in Medicine},
  volume={9},
  number={12},
  pages={1439--1445},
  year={1990}
}

@article{zhou2024,
  author={Zhou, T and Ji, Y},
  title={{On Bayesian sequential clinical trial designs}},
  journal={The New England Journal of Statistics in Data Science},
  volume={2},
  number={1},
  pages={136--151},
  year={2024}
}

@misc{fda2019,
  author={{US Food and Drug Administration}},
  title={{Adaptive Designs for Clinical Trials of Drugs and Biologics: Guidance for Industry}},
  year={2019},
  note={Last accessed 21 April 2026},
  url={https://www.fda.gov/media/78495/download}
}

@misc{fda2010,
  author={{US Food and Drug Administration}},
  title={{Guidance for the Use of Bayesian Statistics in Medical Device Clinical Trials}},
  year={2010},
  note={Last accessed 21 April 2026},
  url={https://www.fda.gov/media/71512/download}
}

@misc{fda2026,
  author={{US Food and Drug Administration}},
  title={{Use of Bayesian Methodology in Clinical Trials of Drug and Biological Products}},
  year={2026},
  note={Draft guidance for industry. Last accessed 21 April 2026},
  url={https://www.fda.gov/media/190505/download}
}

@book{spiegelhalter2004,
  author={Spiegelhalter, D J and Abrams, K R and Myles, J P},
  title={{Bayesian Approaches to Clinical Trials and Health-care Evaluation}},
  publisher={Wiley},
  address={New York},
  year={2004}
}

@book{berry2010,
  author={Berry, S M and Carlin, B P and Lee, J J and Muller, P},
  title={{Bayesian Adaptive Methods for Clinical Trials}},
  publisher={Chapman \& Hall/CRC},
  address={Boca Raton},
  year={2010}
}

@article{gsponer2014,
  author={Gsponer, T and Gerber, F and Bornkamp, B and Ohlssen, D and Vandemeulebroecke, M and others},
  title={{A practical guide to Bayesian group sequential designs}},
  journal={Pharmaceutical Statistics},
  volume={13},
  number={1},
  pages={71--80},
  year={2014}
}

@article{saville2014,
  author={Saville, B R and Connor, J T and Ayers, G D and Alvarez, J},
  title={{The utility of Bayesian predictive probabilities for interim monitoring of clinical trials}},
  journal={Clinical Trials},
  volume={11},
  number={4},
  pages={485--493},
  year={2014}
}

@article{lee2024,
  author={Lee, S Y},
  title={{Using Bayesian statistics in confirmatory clinical trials in the regulatory setting: a tutorial review}},
  journal={BMC Medical Research Methodology},
  volume={24},
  number={1},
  pages={110},
  year={2024}
}

@article{schmidli2014,
  author={Schmidli, H and Gsteiger, S and Roychoudhury, S and O'Hagan, A and Spiegelhalter, D and others},
  title={{Robust meta-analytic-predictive priors in clinical trials with historical control information}},
  journal={Biometrics},
  volume={70},
  number={4},
  pages={1023--1032},
  year={2014}
}

@article{ibrahim2015,
  author={Ibrahim, J G and Chen, M-H and Gwon, Y and Chen, F},
  title={{The power prior: theory and applications}},
  journal={Statistics in Medicine},
  volume={34},
  number={28},
  pages={3724--3749},
  year={2015}
}

@article{jiang2023,
  author={Jiang, L and Nie, L and Yuan, Y},
  title={{Elastic priors to dynamically borrow information from historical data in clinical trials}},
  journal={Biometrics},
  volume={79},
  number={1},
  pages={49--60},
  year={2023}
}

@misc{fda2023,
  author={{US Food and Drug Administration}},
  title={{Considerations for the Design and Conduct of Externally Controlled Trials for Drug and Biological Products}},
  year={2023},
  note={Last accessed 21 April 2026},
  url={https://www.fda.gov/media/164960/download}
}

@article{freedman1994,
  author={Freedman, L S and Spiegelhalter, D J and Parmar, M K B},
  title={{The what, why and how of Bayesian clinical trials monitoring}},
  journal={Statistics in Medicine},
  volume={13},
  number={13-14},
  pages={1371--1383},
  year={1994}
}

@article{winkler2001,
  author={Winkler, R L},
  title={{Why Bayesian analysis hasn't caught on in healthcare decision making}},
  journal={International Journal of Technology Assessment in Health Care},
  volume={17},
  number={1},
  pages={56--66},
  year={2001}
}

@article{pibouleau2011,
  author={Pibouleau, L and Chevret, S},
  title={{Bayesian statistical method was underused despite its advantages in the assessment of implantable medical devices}},
  journal={Journal of Clinical Epidemiology},
  volume={64},
  number={3},
  pages={270--279},
  year={2011}
}

@article{brard2017,
  author={Brard, C and Le Teuff, G and Le Deley, M and Hampson, L V},
  title={{Bayesian survival analysis in clinical trials: what methods are used in practice?}},
  journal={Clinical Trials},
  volume={14},
  number={1},
  pages={78--87},
  year={2017}
}

@article{gerber2016,
  author={Gerber, F and Gsponer, T},
  title={{gsbDesign: an R package for evaluating the operating characteristics of a group sequential Bayesian design}},
  journal={Journal of Statistical Software},
  volume={69},
  number={11},
  pages={1--23},
  year={2016}
}

@article{granholm2022,
  author={Granholm, A and Jensen, A K G and Lange, T and Kaas-Hansen, B S},
  title={{adaptr: an R package for simulating and comparing adaptive clinical trials}},
  journal={Journal of Open Source Software},
  volume={7},
  number={72},
  pages={4284},
  year={2022}
}

@article{couturier2024,
  author={Couturier, D-L and Ryan, E G and Puhr, R and Jaki, T and Heritier, S},
  title={{A fast, flexible simulation framework for Bayesian adaptive designs--the R package BATSS}},
  journal={arXiv},
  pages={2410.02050},
  year={2024}
}

@article{rue2009,
  author={Rue, H and Martino, S and Chopin, N},
  title={{Approximate Bayesian inference for latent Gaussian models by using integrated nested Laplace approximations}},
  journal={Journal of the Royal Statistical Society: Series B (Statistical Methodology)},
  volume={71},
  number={2},
  pages={319--392},
  year={2009}
}

@article{dalal1983,
  author={Dalal, S R and Hall, W J},
  title={{Approximating priors by mixtures of natural conjugate priors}},
  journal={Journal of the Royal Statistical Society: Series B (Methodological)},
  volume={45},
  number={2},
  pages={278--286},
  year={1983}
}

@article{kullback1951,
  author={Kullback, S and Leibler, R A},
  title={{On information and sufficiency}},
  journal={The Annals of Mathematical Statistics},
  volume={22},
  number={1},
  pages={79--86},
  year={1951}
}

@article{dempster1977,
  author={Dempster, A P and Laird, N M and Rubin, D B},
  title={{Maximum likelihood from incomplete data via the EM algorithm}},
  journal={Journal of the Royal Statistical Society: Series B (Methodological)},
  volume={39},
  number={1},
  pages={1--38},
  year={1977}
}

@book{tsybakov2009,
  author={Tsybakov, A B},
  title={{Introduction to Nonparametric Estimation}},
  publisher={Springer},
  address={New York},
  year={2009}
}

@book{vandervaart1998,
  author={van der Vaart, A W},
  title={{Asymptotic Statistics}},
  publisher={Cambridge University Press},
  address={Cambridge},
  year={1998}
}

@article{li2023,
  author={Li, W and Cornelius, V and Finfer, S and Venkatesh, B and Billot, L},
  title={{Adaptive designs in critical care trials: a simulation study}},
  journal={BMC Medical Research Methodology},
  volume={23},
  number={1},
  pages={236},
  year={2023}
}

@article{venkatesh2018,
  author={Venkatesh, B and Finfer, S and Cohen, J and Rajbhandari, D and Arabi, Y and others},
  title={{Adjunctive glucocorticoid therapy in patients with septic shock}},
  journal={New England Journal of Medicine},
  volume={378},
  number={9},
  pages={797--808},
  year={2018}
}

@manual{granholm2026,
  author={Granholm, A and Kaas-Hansen, B S and Jensen, A K G and Lange, T},
  title={{adaptr: Adaptive Trial Simulator}},
  year={2026},
  note={{R package version 1.5.0}},
  url={https://cran.r-project.org/web/packages/adaptr}
}

@manual{couturier2025,
  author={Couturier, D-L and Ryan, E G and Puhr, R and Jaki, T and Heritier, S},
  title={{BATSS: Bayesian Adaptive Trial Simulator Software (BATSS) for Generalised Linear Models}},
  year={2025},
  note={{R package version 1.1.1}},
  url={https://cran.r-project.org/web/packages/BATSS}
}

@article{ferkingstad2015,
  author={Ferkingstad, E and Rue, H},
  title={{Improving the INLA approach for approximate Bayesian inference for latent Gaussian models}},
  journal={Electronic Journal of Statistics},
  volume={9},
  number={2},
  pages={2706--2731},
  year={2015}
}

@manual{weber2026,
  author={Weber, S and Neuenschwander, B and Schmidli, H and Magnusson, B and Li, Y and others},
  title={{RBesT: R Bayesian Evidence Synthesis Tools}},
  year={2026},
  note={R package version 1.9-0},
  url={https://CRAN.R-project.org/package=RBesT}
}

@book{law2015,
  author={Law, Averill M.},
  title={{Simulation Modeling and Analysis}},
  edition={5},
  publisher={McGraw-Hill Education},
  address={New York},
  year={2015}
}

@article{lewis1994,
  author={Lewis, R J and Berry, D A},
  title={{Group sequential clinical trials: a classical evaluation of Bayesian decision-theoretic designs}},
  journal={Journal of the American Statistical Association},
  volume={89},
  number={428},
  pages={1528--1534},
  year={1994}
}

@article{broglio2022,
  author={Broglio, K and Meurer, W J and Durkalski, V and Pauls, Q and Connor, J and others},
  title={{Comparison of Bayesian vs frequentist adaptive trial design in the stroke hyperglycemia insulin network effort trial}},
  journal={JAMA Network Open},
  volume={5},
  number={5},
  pages={e2211616},
  year={2022}
}

@article{he2025,
  author={He, Z and Cro, S and Billot, L},
  title={{Optimal scheduling of interim analyses in group sequential trials}},
  journal={arXiv},
  pages={2509.05537},
  year={2025}
}

\clearpage
\appendix
\setcounter{figure}{0}
\setcounter{table}{0}

\section{Approximation diagnostics}
\label{sec:A}

\subsection{Prior approximation diagnostics for the non-conjugate prior}
\label{sec:A1}
The proposed framework replaces a user-specified prior $\pi_{0}(\vartheta)$ with the Beta-mixture approximation $\hat{\pi}_{0}^{L}(\vartheta)$ of Section~\ref{sec:23}, selected through the empirical forward KL within-tolerance rule on a set of independent draws from $\pi_{0}(\vartheta)$. Before using $\hat{\pi}_{0}^{L}$ in trial simulation, we recommend reporting four diagnostics that examine the quality of the approximation: (i) the forward KL divergence $D_{L}=D_{KL}(\pi_{0}\parallel\hat{\pi}_{0}^{L})$ as a function of the number of mixture components $L$; (ii) a density overlay of $\pi_{0}(\vartheta)$ and $\hat{\pi}_{0}^{L}(\vartheta)$ on the response-rate scale; (iii) prior tail probabilities at the decision-rule thresholds, namely $\mathbb{P}_{\hat{\pi}_{0}^{L}}(\Delta<0)$ and $\mathbb{P}_{\hat{\pi}_{0}^{L}}(\Delta<\delta)$, evaluated by independent prior Monte Carlo draws and compared with the same quantities under $\pi_{0}$; and (iv) the sensitivity of these summaries to $L$. The first three diagnostics establish that $\hat{\pi}_{0}^{L}$ matches $\pi_{0}$ at the resolution that the decision rules actually use, and the fourth confirms that the within-tolerance rule has selected a parsimonious mixture beyond which additional components inflate the prior effective sample size without improving fit.

\begin{figure}[!ht]
	\centering
	\includegraphics[width=\linewidth]{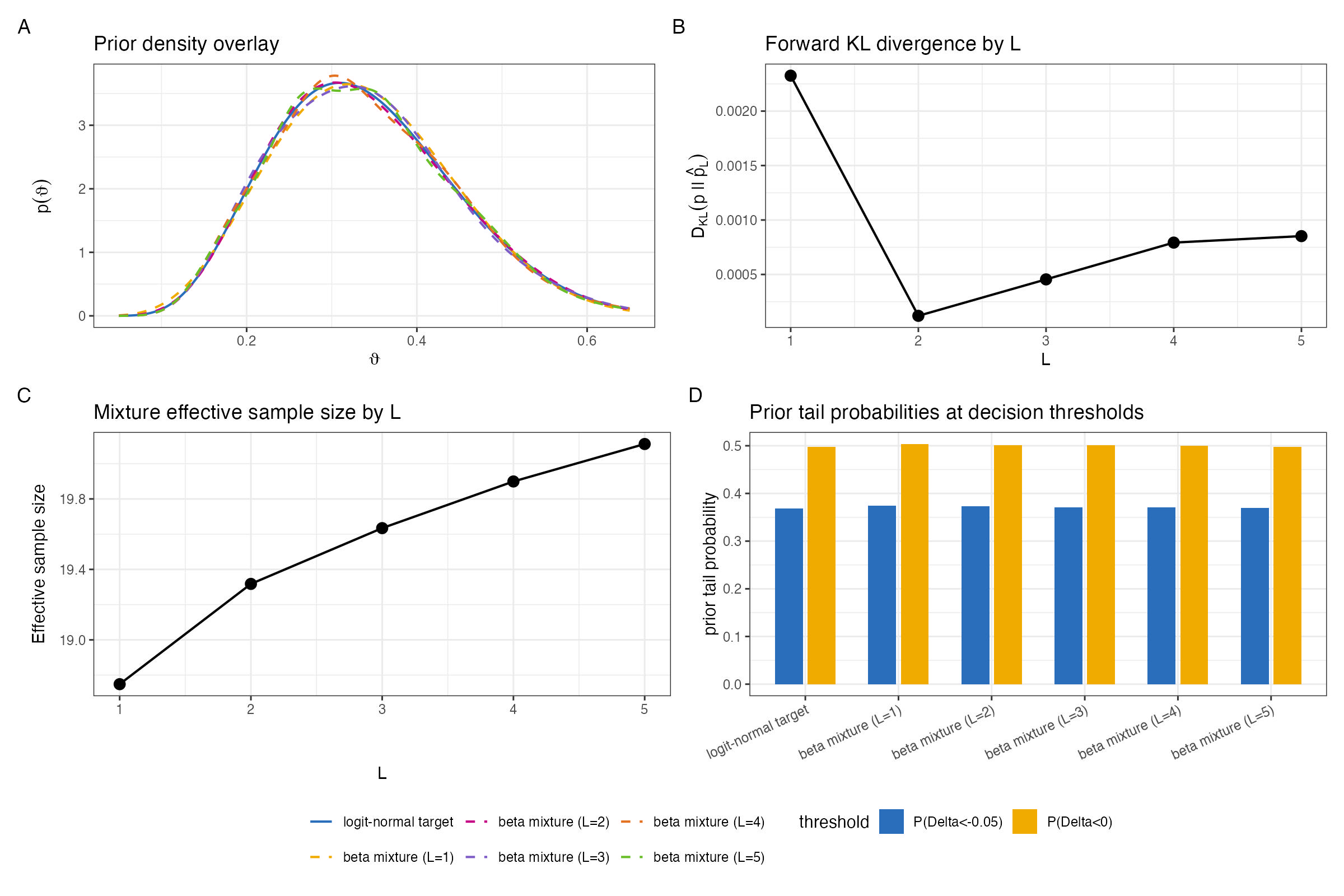}
	\caption{Prior approximation diagnostics for the logit-normal example of Section~\ref{sec:332}. Panel A: density overlay of the target logit-normal prior on the response-rate scale and its Beta-mixture approximations at $L\in\{1,2,\ldots,5\}$. Panel B: forward KL divergence $D_{L}$ as a function of $L$. Panel C: effective sample size of the fitted mixture as a function of $L$. Panel D: prior tail probabilities at the decision-rule thresholds, $\mathbb{P}(\Delta<0)$ and $\mathbb{P}(\Delta<-0.05)$, evaluated under the target logit-normal prior and under each fitted Beta mixture by $50{,}000$ independent prior Monte Carlo draws per arm.}
	\label{fig:A11}
\end{figure}

We illustrate the four diagnostics on the logit-normal example of Section~\ref{sec:332}. Figure~\ref{fig:A11} shows the density overlay (panel A), the KL divergence by $L$ (panel B), the effective sample size of the fitted mixture by $L$ (panel C), and the prior tail probabilities at the decision thresholds, evaluated under the target prior and under each fitted mixture (panel D). Table~\ref{tab:A11} reports the numerical values. 

The KL divergence falls by more than an order of magnitude between $L=1$ and $L=2$ and remains at the same order of magnitude (around $10^{-4}$) for $L\in\{2,\ldots,5\}$, with $L=2$ giving the lowest empirical KL. The mild non-monotonicity at $L \geq 2$ reflects the EM algorithm converging to different local optima as the number of components increases. Applying the selection rule of Section~\ref{sec:23} with $\varepsilon=10^{-3}$ to these values gives a minimum forward KL of $D_{\min}\approx 1.2\times10^{-4}$ at $L=2$: the $L=1$ fit lies about $2.2\times10^{-3}$ above $D_{\min}$ and is rejected, whereas $L=2$ attains the minimum and is therefore selected, so $\hat{L}=2$.

The Section~\ref{sec:332} benchmark uses the same per-$L$ fitting procedure, so the mixture it employs is the $\hat{L}=2$ fit reported here. Mixtures with $L\geq 2$ have effective sample sizes within about one patient of each other, so additional components add negligible prior information beyond $\hat{L}=2$. Every fitted mixture's tail probabilities at the decision thresholds lie within $0.006$ of the target prior across all $L\geq 1$, confirming that the approximation is faithful at the resolution at which the design decisions are taken. The fitted $\hat{L}=2$ mixture is therefore both adequate and parsimonious. Because agreement at the prior level need not contract uniformly to the calibrated operating characteristics, the prior-tail diagnostic of Table~\ref{tab:A11} is necessary but not sufficient. We therefore report the stronger, operating-characteristic-level diagnostic directly. Table~\ref{tab:A12} re-evaluates the calibrated three-look design from Section~\ref{sec:35} under the logit-normal prior approximated with $L\in\{1,2,3\}$. Simulated event counts are held fixed across $L$, so any differences arise solely from the prior approximation. Across $L\in\{1,2,3\}$, the type I error rate varies by only about $0.01$ percentage points across the displayed values ($2.33\%$ to $2.34\%$), and by $0.005$ percentage points on the unrounded estimates, below the $0.016$ percentage-point binomial Monte Carlo SE at $R=10^{6}$. The power varies by $0.020$ percentage points (around $91.1\%$), and the expected sample sizes differ by at most one patient under $H_{0}$ and three under $H_{1}$. Thus, the within-tolerance choice $\hat{L}=2$ yields operating characteristics indistinguishable from those for $L=1$ and $L=3$, the design remains within the $\alpha\leq 2.5\%$ target for all three approximations, and the prior-tail agreement of Table~\ref{tab:A11} propagates to the calibrated operating characteristics for this boundary design. Nevertheless, we recommend reporting the operating-characteristic-level check of Table~\ref{tab:A12} as standard practice for any non-conjugate prior calibration, since small shifts in the posterior tail at the decision threshold can in principle flip stopping decisions for a design poised on a regulatory boundary.

\begin{table}[!ht]
	\centering
	\begin{tabular}{rrrrr}
		\toprule
		$L$ & $D_{L}$ & ESS & $\mathbb{P}(\Delta<0)$ & $\mathbb{P}(\Delta<-0.05)$ \\
		\midrule
		Target & \textemdash & \textemdash & $0.498$ & $0.369$ \\
		$1$ & $2.33 \times 10^{-3}$ & $18.7$ & $0.504$ & $0.374$ \\
		$2$ & $1.20 \times 10^{-4}$ & $19.3$ & $0.501$ & $0.373$ \\
		$3$ & $4.55 \times 10^{-4}$ & $19.6$ & $0.501$ & $0.371$ \\
		$4$ & $7.93 \times 10^{-4}$ & $19.9$ & $0.500$ & $0.371$ \\
		$5$ & $8.52 \times 10^{-4}$ & $20.1$ & $0.497$ & $0.370$ \\
		\bottomrule
	\end{tabular}
	\caption{Prior approximation diagnostics for the logit-normal example of Section~\ref{sec:332}. ``$D_{L}$'' is the empirical forward KL divergence of the fitted mixture $\hat{\pi}_{0}^{L}$ from the target prior $\pi_{0}$, estimated on $50{,}000$ independent draws. ``ESS'' is the effective sample size of the fitted Beta mixture, reported by \texttt{RBesT::ess} under the expected local-information-ratio criterion. The tail probabilities $\mathbb{P}(\Delta<0)$ and $\mathbb{P}(\Delta<-0.05)$ are estimated from $50{,}000$ independent draws of $\Delta=\vartheta_{t}-\vartheta_{c}$ under independent per-arm priors. The ``Target'' row reports the target logit-normal prior, and the remaining rows report the Beta-mixture approximation at the indicated $L$.}
	\label{tab:A11}
\end{table}

\begin{table}[!ht]
	\centering
	\begin{tabular}{ccccc}
		\toprule
		$L$ & Type I error & Power & $\mathbb{E}(N\mid H_{0})$ & $\mathbb{E}(N\mid H_{1})$ \\
		\midrule
		$1$            & $2.34\%$ & $91.08\%$ & $3{,}485$ & $3{,}186$ \\
		$2$ ($\hat{L}$) & $2.34\%$ & $91.08\%$ & $3{,}484$ & $3{,}186$ \\
		$3$            & $2.33\%$ & $91.10\%$ & $3{,}484$ & $3{,}183$ \\
		\bottomrule
	\end{tabular}
	\caption{Operating-characteristic sensitivity to the Beta-mixture order $L$ for the Section~\ref{sec:35} calibrated three-look design ($K=3$, $\boldsymbol{\gamma}_{1:3}^{E}=(0.999,0.999,0.977)$, $\gamma^{F}=0.90$, binding futility), under the informative logit-normal prior of Section~\ref{sec:332} approximated by a Beta mixture of order $L$. Each row uses $R=10^{6}$ virtual trials with a common seed, so the simulated event counts are identical across $L$ and only the prior approximation varies. The within-tolerance rule of Section~\ref{sec:23} selects $\hat{L}=2$. Across $L\in\{1,2,3\}$ the type I error rate spans $0.005$ percentage points (below the $0.016$ percentage-point binomial Monte Carlo SE at $R=10^{6}$) and the power spans $0.020$ percentage points, confirming that the prior-tail agreement of Table~\ref{tab:A11} propagates to operating characteristics that are, for design purposes, invariant to $L$ at and beyond $\hat{L}$.}
	\label{tab:A12}
\end{table}

In practice, we recommend reporting Figure~\ref{fig:A11} and Table~\ref{tab:A11} together with the operating characteristics for any non-conjugate prior. If the within-tolerance selected $\hat{L}$ lies above the desired computational budget, the tail-probability diagnostic provides a practical fallback. Specifically, if the target and approximate tail posterior probabilities at the decision thresholds agree within the Monte Carlo SE of the simulation budget adopted to evaluate the operating characteristics, the approximation is adequate for the decision-rule calibration.

\subsection{Quadrature convergence diagnostics}
\label{sec:A2}
The closed-form per-look posterior tail probability of Section~\ref{sec:22} reduces to a one-dimensional integral on $(0,1)$, which the framework evaluates by Gauss--Legendre quadrature with $Q$ nodes ($Q=128$ by default). To confirm that this fixed-$Q$ rule is adequate for the regimes considered in this paper, we run a convergence study at $Q\in\{64,128,256,512\}$ under both the $\operatorname{Beta}(1,1)$ baseline of Section~\ref{sec:331} and the logit-normal Beta-mixture of Section~\ref{sec:332}.

\begin{figure}[!ht]
	\centering
	\includegraphics[width=\linewidth]{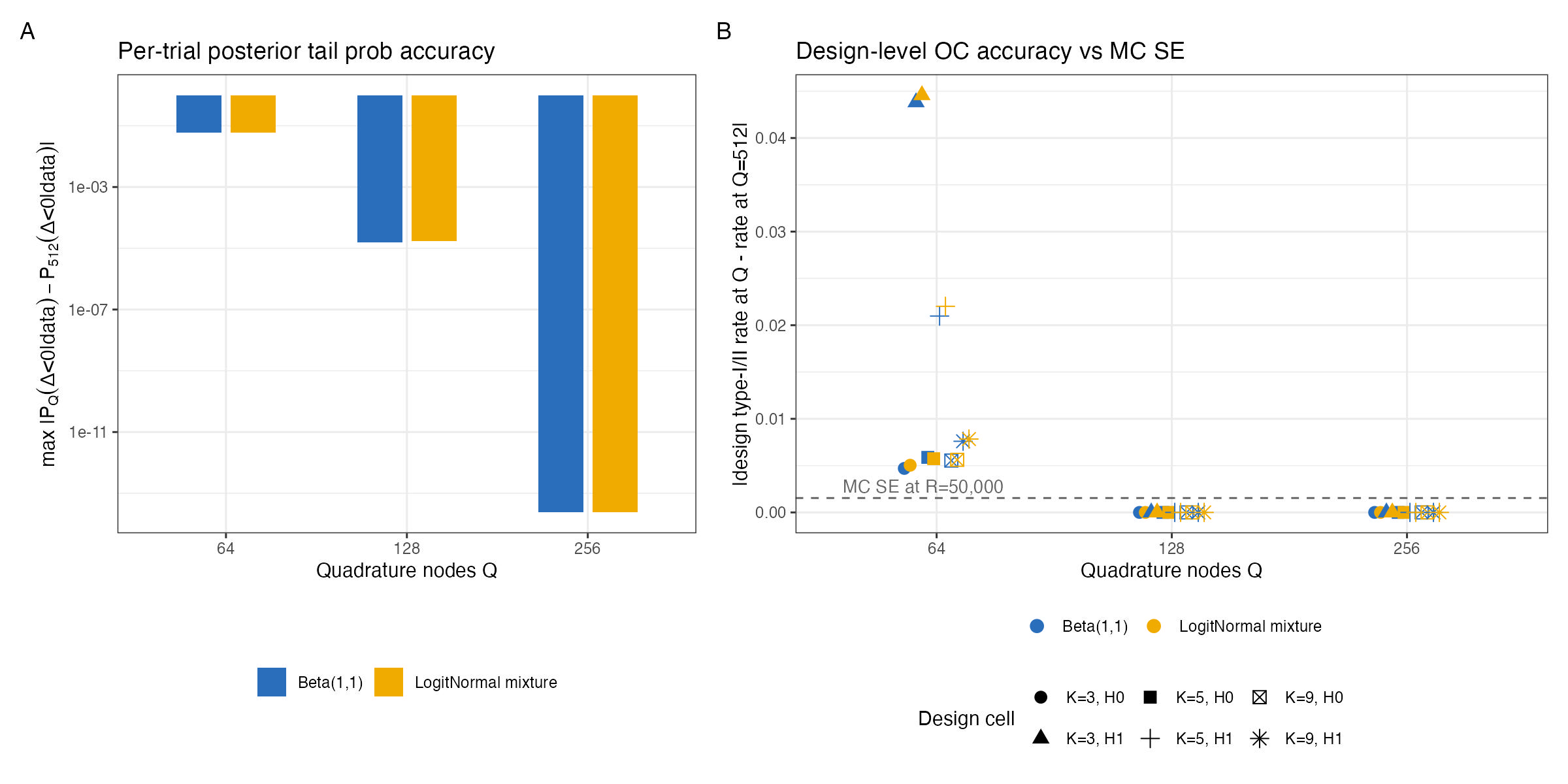}
	\caption{Quadrature convergence of the proposed framework. Panel A: maximum absolute difference between the per-trial posterior tail probability $\mathbb{P}(\Delta<0\mid I_{k},J_{k})$ at $Q$ Gauss--Legendre nodes and the $Q=512$ reference, across a representative grid of $(I_{k},J_{k})$ event-count cells at the early ($k=3$), mid ($k=6$), and final ($k=9$) interim looks of the $K=9$ design at $N=3{,}800$, for both the $\operatorname{Beta}(1,1)$ prior and the logit-normal Beta-mixture prior. Panel B: absolute difference between the design-level type I or type II error rate estimate at $Q$ nodes and the $Q=512$ reference, evaluated at the Section~\ref{sec:33} protocol ($\gamma^{E}=0.99$, $\gamma^{F}=0.90$, $R=50{,}000$ virtual trials) for $K\in\{3,5,9\}$ and under both $H_{0}$ and $H_{1}$. The dashed line denotes the maximum binomial Monte Carlo SE across the cells reported, $\sqrt{p(1-p)/R}\approx 0.15$ pp at $R=50{,}000$, achieved at the largest effective rate in the panel ($p\approx 0.87$, corresponding to the type II error rate cells). The type I error rate cells ($p\approx 0.025$) sit at $\approx 0.07$ pp.}
	\label{fig:A21}
\end{figure}

We compare both (i) the per-trial posterior tail probabilities at a representative set of $(I_{k},J_{k})$ cells spanning the early, mid, and final analyses of the $K=9$ design at $N=3{,}800$ and (ii) the design-level type I and type II error rates at the Section~\ref{sec:33} protocol, using $\gamma^{E}=0.99$, $\gamma^{F}=0.90$, $K\in\{3,5,9\}$, and $R=50{,}000$. Figure~\ref{fig:A21} reports the result against the $Q=512$ reference. 

At $Q=128$ the maximum per-trial absolute difference is well below $10^{-3}$ under either prior, and the design-level type I and type II error rate differences are at most a small fraction of the Monte Carlo SE implied by $R=50{,}000$. Thus, the fixed-$Q$ quadrature introduces deterministic numerical error that is at least an order of magnitude smaller than the simulation noise at the routine evaluation budget. For a more conservative margin, $Q$ can be increased to $256$ at marginal computational cost. The per-trial accuracy improves by several orders of magnitude at each doubling of $Q$, with the per-doubling factor itself increasing in $Q$, consistent with the rapid (spectral-rate) convergence of Gauss--Legendre quadrature on integrands that are smooth on $(0,1)$ and bounded at the endpoints, which is exponential-in-$Q$ when the integrand is analytic (the conjugate $\operatorname{Beta}(1,1)$ case) and remains effectively super-algebraic for the smooth but non-analytic logit-normal Beta-mixture posteriors exercised in Section~\ref{sec:332}.

The convergence reported here is established for the regime exercised by the ADRENAL worked example, with per-arm sample sizes up to about $1{,}900$ and event rates near $\vartheta_{c}=0.33$, where the per-look Beta and Beta-mixture posteriors are only moderately concentrated (posterior SD around $0.011$ at the final analysis). Since the fixed-$Q$ Gauss--Legendre rule distributes its nodes across the whole of $(0,1)$ by a single global rule, the accuracy it attains at a given $Q$ decreases as the posterior concentrates into a narrow subinterval. Two settings sharpen the posterior in this way. The first is a rare-event endpoint, where $\vartheta_{c}$ lies close to $0$ or $1$ and the Beta posterior concentrates near a boundary of $(0,1)$. The second is a trial substantially larger than ADRENAL, where the posterior SD falls below the values seen here. The spectral-rate argument above guarantees that the deterministic error still decays rapidly in $Q$, but it does not fix the $Q$ at which a target accuracy is reached, and that $Q$ grows as the posterior sharpens. Outside the regime validated here, we recommend that users repeat the convergence check of this appendix at their own sample sizes and event rates, raising $Q$ until the per-trial tail probabilities and the design-level error rates stabilise. An alternative is a quadrature rule that concentrates nodes near the posterior mode, such as a logit change of variables or an adaptive peak-centred rule, which keeps a fixed node count adequate when the posterior is sharply concentrated.

\end{document}